\documentclass[aps,pra,reprint,superscriptaddress,floatfix]{revtex4-1}
\pdfoutput=1

\usepackage{graphicx}
\usepackage{amsmath}
\usepackage{natbib}
\usepackage[usenames,dvipsnames,svgnames,table]{xcolor}
\usepackage{upgreek}
\usepackage{booktabs}
\usepackage{siunitx}
\usepackage{multirow}
\usepackage{physics}
\usepackage{soul}

\usepackage{hyperref}
\hypersetup{
    colorlinks = true,
    citecolor = {blue},
    linkcolor = {blue},
    urlcolor = {blue}
}

\DeclareSIUnit{\belmilliwatt}{Bm}
\DeclareSIUnit{\dBm}{\deci\belmilliwatt}

\AtBeginDocument{
\heavyrulewidth=.08em
\lightrulewidth=.05em
\cmidrulewidth=.03em
\belowrulesep=.65ex
\belowbottomsep=0pt
\aboverulesep=.4ex
\abovetopsep=0pt
\cmidrulesep=\doublerulesep
\cmidrulekern=.5em
\defaultaddspace=.5em
}

\renewcommand{\ket}[1]{\vert \text{#1}\rangle}
\newcommand{\ketit}[1]{\vert #1 \rangle}

\newcommand{\pgg}{\mathcal{P}_\textrm{gg}}
\newcommand{\pge}{\mathcal{P}_\textrm{ge}}
\newcommand{\peg}{\mathcal{P}_\textrm{eg}}
\newcommand{\pee}{\mathcal{P}_\textrm{ee}}
\newcommand{\tg}{\tau_\textnormal{g}}

\graphicspath{ {./images/} }

\begin{document}

\title{Implementation of a generalized \textsf{CNOT} gate between fixed-frequency transmons}

\author{Shavindra P. Premaratne}
\email{shavi@umd.edu}
\affiliation{Department of Physics, University of Maryland, College Park, Maryland 20742, USA}
\affiliation{Laboratory for Physical Sciences, College Park, Maryland 20740, USA}
\author{Jen-Hao Yeh}
\affiliation{Department of Physics, University of Maryland, College Park, Maryland 20742, USA}
\affiliation{Laboratory for Physical Sciences, College Park, Maryland 20740, USA}
\author{F. C. Wellstood}
\affiliation{Department of Physics, University of Maryland, College Park, Maryland 20742, USA}
\affiliation{Joint Quantum Institute and Center for Nanophysics and Advanced Materials, College Park, Maryland 20742, USA}
\author{B. S. Palmer}
\affiliation{Laboratory for Physical Sciences, College Park, Maryland 20740, USA}

\date{\today}

\begin{abstract}
\noindent 
We have embedded two fixed-frequency Al/AlO$_{\textrm{x}}$/Al transmons, with ground-to-excited transition frequencies at \SI{6.0714}{\GHz} and \SI{6.7543}{\GHz}, in a single 3D Al cavity with a fundamental mode at \SI{7.7463}{\GHz}. Strong coupling between the cavity and each transmon results in an effective qubit-qubit coupling strength of \SI{26}{\MHz} and a \SI{-1}{\MHz} dispersive shift in each qubit's transition frequency, depending on the state of the other qubit. Using the all-microwave \textsf{SWIPHT} (Speeding up Waveforms by Inducing Phases to Harmful Transitions) technique \citep{Economou}, we demonstrate the operation of a generalized controlled-not (\textsf{CNOT}) gate between the two qubits, with a gate time $\tg=\SI{907}{\ns}$ optimized for this device. Using quantum process tomography we find that the gate fidelity is 83\%--84\%, somewhat less than the 87\% fidelity expected from relaxation and dephasing in the transmons during the gate time.
\end{abstract}

\maketitle

\section{Introduction \label{sec:intro}}

For a quantum computer to solve classically-hard problems such as prime factorization \citep{Shor} and quantum simulations \citep{Georgescu}, it must be able to produce arbitrary single qubit states and have a quantum gate capable of entangling two or more qubits \citep{Nielsen2014, Barenco}. Many different entangling gates have been implemented in superconducting qubits \citep{Gambetta}, trapped ions \citep{Monroe}, semiconductor quantum dots \citep{Zajac}, and nitrogen-vacancy centers in diamond \citep{Wei}. Understanding the trade-offs and performance of gates for quantum computing applications is an active area of research, with scalability, ease of use, gate speed and gate fidelity being important factors in determining the suitability of a gate for a particular quantum system. 

For superconducting qubits, entangling gates can be broadly classified into those based on flux-tunability or other parametric control \citep{McKay2016,Reagor} and those based on all-microwave control \citep{Gambetta}. In circuit quantum electrodynamics (cQED) \citep{Blais2004} based systems, flux-tunable elements have enabled fast and efficient gates at the expense of more control lines and potentially smaller coherence times from operating off the sweet spot. Examples of such gates include the dynamically tuned $\textsf{C-Phase}$ gate \citep{Strauch2003, Barends2014, DiCarlo2009}, and the direct-resonance $\sqrt{\textsf{iSWAP}}$ gate \citep{Bialczak2010, Dewes2012}. Gates utilizing all-microwave controls are typically slower but achieve longer qubit coherence times. Examples include the cross-resonance (CR) gate \citep{Rigetti2010, Chow2011}, resonator-sideband-induced $\sqrt{\textsf{iSWAP}}$ gate \citep{Leek2009}, the microwave-activated $\textsf{C-Phase}$ gate \citep{Chow2013}, the resonator-induced-phase (RIP) gate \citep{Cross2015, Paik2016}, and a $\textsf{CNOT}$ gate between two encoded multiphoton qubits in 3D cavities \citep{Rosenblum2018}. A limitation of the CR gate is the requirement that the qubit transition frequencies must be within a relatively narrow frequency window. This not only makes device fabrication more challenging but leads to frequency crowding issues when scaling to many qubits \citep{Schutjens2013, Vesterinen2014}. The CR gate also requires that each qubit be individually addressable. In contrast, the RIP gate can operate with a common drive using a bus cavity and a relatively large frequency detuning between qubits. As the interaction between the qubits is directly mediated through the cavity, a high-quality-factor bus cavity is required to prevent decoherence; but this can limit the readout speed if the cavity is used for qubit readout.

Here we describe the implementation of a recently proposed all-microwave two-qubit entangling gate known as ``Speeding up Waveforms by Inducing Phases to Harmful Transitions'' or \textsf{SWIPHT} \citep{Economou,Deng}. $\textsf{SWIPHT}$ is a generalized $\textsf{CNOT}$ gate that yields a $\pi$-rotation (bit-flip) on the desired transition and a $2\pi$-rotation (no bit-flip) on an undesired transition. In the $\textsf{SWIPHT}$ technique, the resonator plays a subsidiary role, allowing us to use a relatively low-quality-factor cavity to achieve a fast qubit readout. In Sec.~\ref{sec:expr_setup} we describe the Hamiltonian for the system, the pulse shaping required for the gate and the joint-readout technique we used. Sec.~\ref{sec:qst} describes how state tomography was used to verify that the proper states were being generated in the two-qubit system. Sec.~\ref{sec:qpt} discusses how process tomography was used to characterize the fidelity and purity of the \textsf{SWIPHT} gate. Additional results from process tomography on other gates, including modified $\textsf{SWIPHT}$ gates, are also presented. Finally in Sec.~\ref{sec:conclusions}, we conclude with a discussion of the limitations of the present implementation and suggest how the gate time and fidelity could be improved.


\section{Experimental Setup \label{sec:expr_setup}}

\begin{figure}[t]
\centering
\includegraphics[width=1\columnwidth]{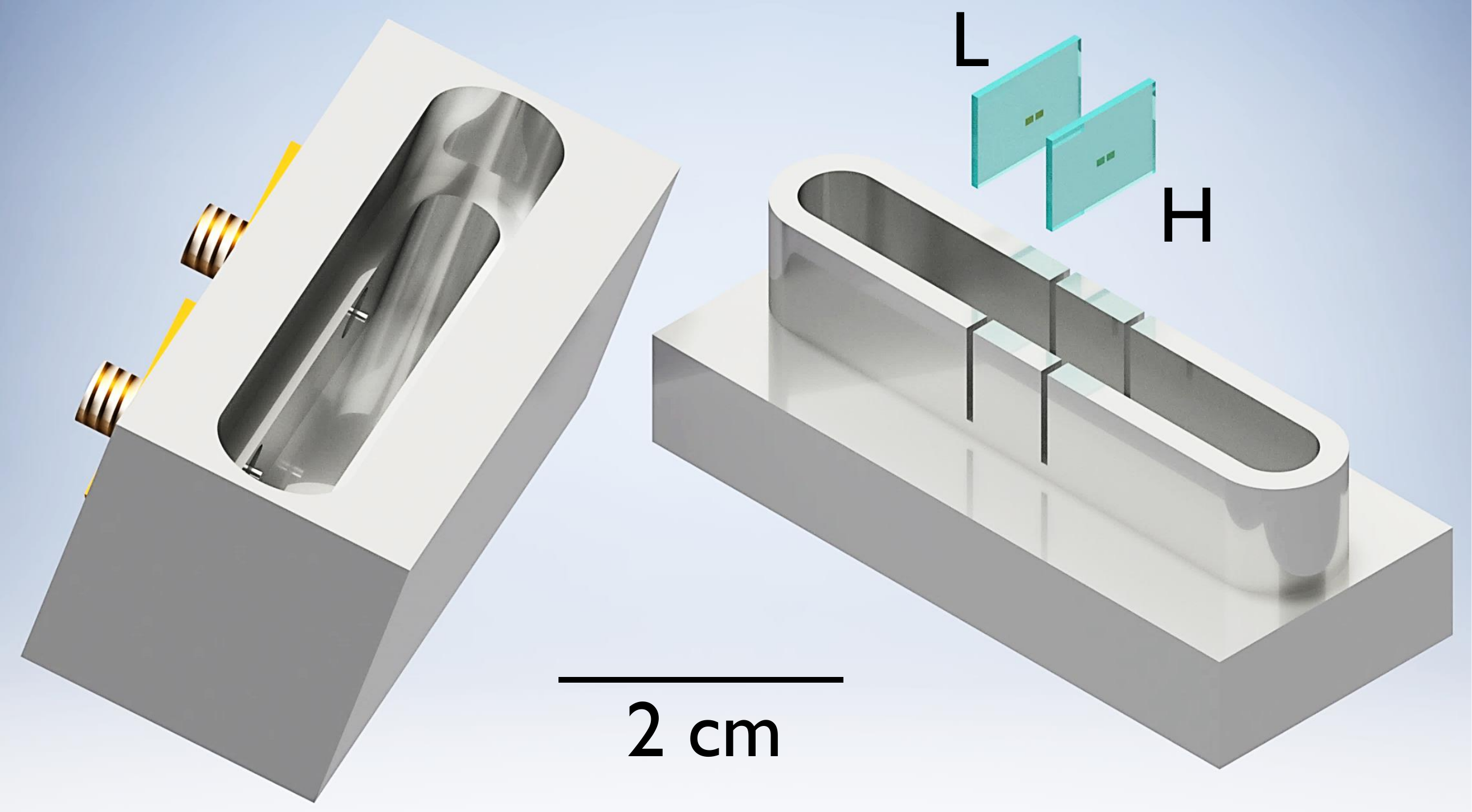}
\caption{Computer generated rendering of the transmon qubit chips (L and H), and the two sections of the 3D Al cavity. \label{fig:TransCav}}
\end{figure}

Our cQED system consisted of two transmons \citep{Koch} on separate sapphire substrates that were mounted in a single superconducting Al microwave cavity (see Fig.~\ref{fig:TransCav}) \citep{Paik2011}. The transition frequencies between the two lowest levels ($\ket{g}$ and $\ket{e}$) of the transmons were $\omega_\textrm{L}/2\pi = \SI{6.07135}{\GHz}$ for qubit L ($Q_\textnormal{L}$) and $\omega_\textrm{H}/2\pi = \SI{6.75427}{\GHz}$ for qubit H ($Q_\textnormal{H}$). The direct capacitive coupling between the transmons, combined with the indirect coupling through the \SI{7.7463}{\GHz} cavity resonance, resulted in a qubit-qubit dispersive shift \citep{Majer2007} of $2\chi_\textrm{qq}/2\pi = \SI{-1.04}{\MHz}$ (see Appendix~\ref{app:hamiltonian} for detailed system parameters). While this always-on interaction leads to a two-qubit phase gate of the form $\frac{1}{\sqrt{2}} \left( \hat{I}\hat{Z}-i\hat{Z}\hat{I}\right)$ in a time $\pi/2\chi_\textnormal{qq}$, we implement a driven generalized \textsf{CNOT} gate here. 

Since our gates do not involve occupation of excited states of the cavity or transmon levels higher than $\ket{e}$ (\textit{i.e.} $\ket{f},~\ket{h},\dots$), for simplicity, we restrict consideration to the computational two-qubit subspace of the system spanned by $\left\{ \ket{gg}, \ket{ge}, \ket{eg}, \ket{ee} \right\}$, where the labels within kets represent the states of $Q_\textnormal{L}$ and $Q_\textnormal{H}$ respectively. In this case, the undriven system can be described by the Hamiltonian:
\begin{align}
\mathcal{H}_0 &= \frac{\hbar}{2} \left(\omega_\textrm{L} + \chi_\textnormal{qq} \right) \sigma_z^\textnormal{(L)} + \frac{\hbar}{2} \left(\omega_\textrm{H} + \chi_\textnormal{qq} \right) \sigma_z^\textnormal{(H)} \nonumber \\
&\hspace{2em} + \frac{\hbar \chi_\textrm{qq}}{2} \sigma_z^\textnormal{(L)} \sigma_z^\textnormal{(H)}, \label{eq:H0}
\end{align}
where $\sigma_z^\textnormal{(L)}$ and $\sigma_z^\textnormal{(H)}$ are the Pauli-Z operators for the qubits L and H respectively. 

The two-qubit system was driven according to the $\textsf{SWIPHT}$ protocol, utilizing qubit L as the control and qubit H as the target for the generalized $\textsf{CNOT}$ gate \citep{Economou}. The drive Hamiltonian can be expressed as:
\begin{align}
\mathcal{H}_\textrm{d} &= \frac{\hbar \Omega(t)}{2} \left( \sigma_\textrm{H}^- \, e^{i(\omega_\textrm{d} t + \phi_\textrm{d})} + \sigma_\textrm{H}^+ \, e^{-i(\omega_\textrm{d} t + \phi_\textrm{d})} \right),
\label{eq:Hdrive}
\end{align}
where $\Omega(t)$ is the pulse shape or envelope of the microwave drive signal (expressed as the Rabi frequency it produces), $\sigma_\textrm{H}^\pm$ is the raising ($+$) or lowering ($-$) operator for qubit H, $\phi_\textrm{d}$ is the drive phase and $\omega_\textrm{d}$ is the drive frequency for the gate pulse. Due to the relatively large detuning between the qubits ($\sim \SI{700}{\MHz}$), we have neglected the off-resonant drive term on the control qubit. The full system Hamiltonian for the two-qubit subspace is then given by,
\begin{align}
\mathcal{H} = \mathcal{H}_0 + \mathcal{H}_\textrm{d}.
\label{eq:Htotal}
\end{align}

\subsection{Pulse Shaping \label{ssec:pulse_shaping}}

The $\textsf{SWIPHT}$ protocol was implemented by driving the two-qubit system at frequency $\omega_\textrm{d} = \omega_\textrm{H}$ using a specific analytically derived pulse shape $\Omega(t)$ that depends only on $\chi_\textnormal{qq}$ \citep{Economou}. The duration of the pulse is
\begin{align}
\tg = \frac{5.87}{2  \left| \chi_\textrm{qq} \right|}. \label{eq:SWIPHT_length}
\end{align}
and the pulse shape can be written as
\begin{align}
\Omega(t) &= \frac{\ddot{\gamma}}{\sqrt{\chi_\textrm{qq}^2 - \dot{\gamma}^2}} - 2\sqrt{\chi_\textrm{qq}^2 - \dot{\gamma}^2} \cot{(2\gamma)},
\end{align}
where
\begin{align}
\gamma(t)  &= 138.9 \left(\frac{t}{\tg} \right)^4 \left(1 - \frac{t}{\tg} \right)^4 + \frac{\pi}{4}. \label{eq:xi_t}
\end{align}
From Eqs.~\ref{eq:SWIPHT_length}--\ref{eq:xi_t} it can be shown that the maximum amplitude of the pulse is
\begin{align}
\Omega_\textrm{max} = 0.887 \times 2  \left| \chi_\textrm{qq} \right|. \label{eq:max_amplitude}
\end{align}

With this choice for the pulse shape, the operation of the \textsf{SWIPHT} gate can be written in the two-qubit basis $\left\{ \ket{gg}, \ket{ge}, \ket{eg}, \ket{ee} \right\}$ as 
\begin{align}
\textsf{SWIPHT} = \left(
	\begin{array}{cccc}
		0 & e^{i \phi_\textrm{d}} & 0 & 0 \\
		e^{-i \phi_\textrm{d}} & 0 & 0 & 0 \\
		0 & 0 & e^{i \xi} & 0 \\
		0 & 0 & 0 & e^{i \zeta}
	\end{array}
\right) \label{eq:gen_CNOT}
\end{align}
with $\phi_\textnormal{d}=0$, $\xi \approx 1.16~\textrm{rad}$, and $\zeta \approx 1.98~\textrm{rad}$ (note $\xi + \zeta = \pi$). As with a standard \textsf{CNOT} operation, the target qubit is flipped depending on the control qubit. However, this is a generalized \textsf{CNOT} gate since the control qubit also acquires extra $z$-rotations (represented by $\xi$ and $\zeta$) due to the non-trivial $2\pi$-rotation imposed on the harmful transition as discussed next.

\begin{figure}[b]
\centering
\includegraphics[width=0.8\columnwidth]{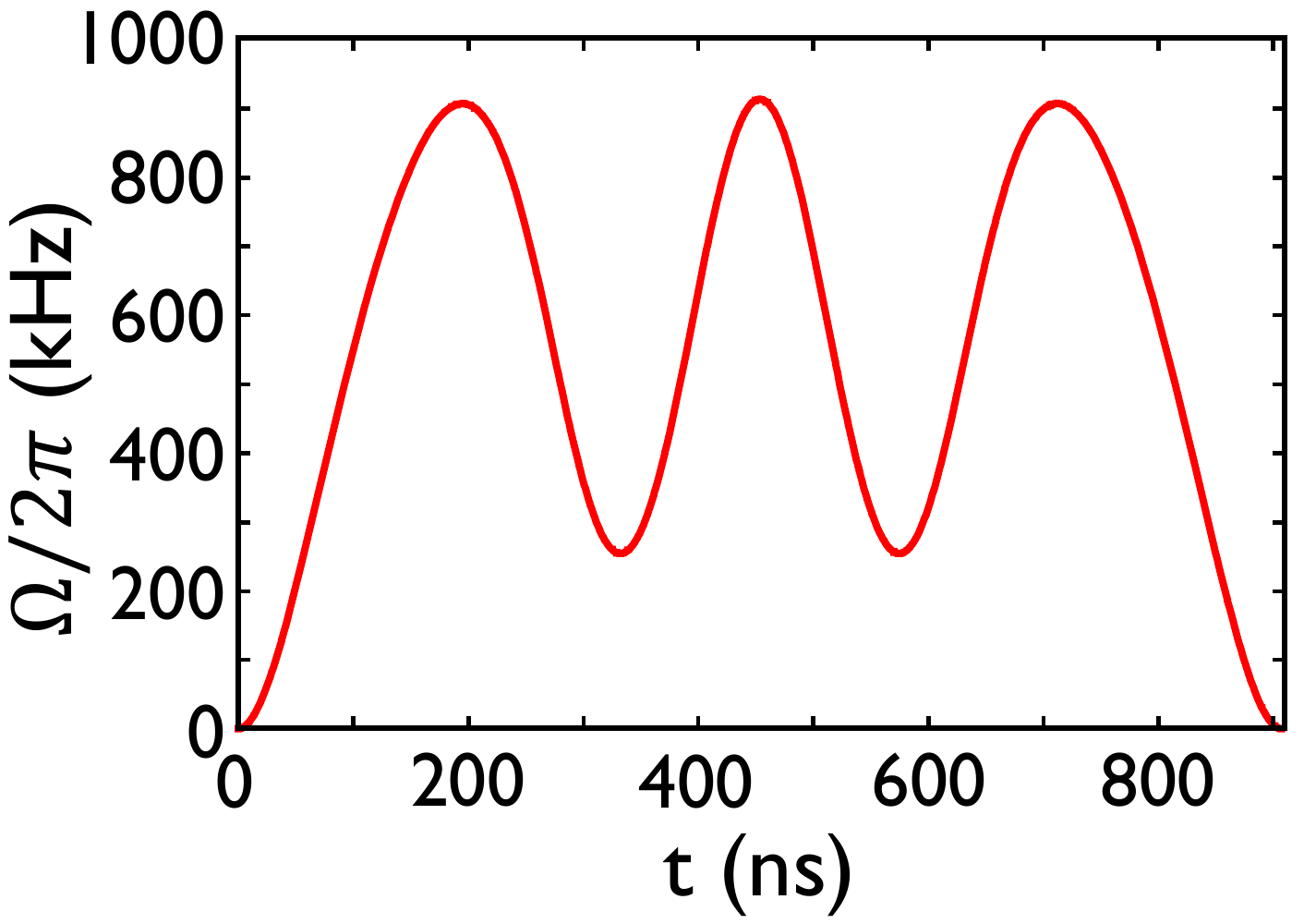}
\caption{Plot of analytically derived envelope $\Omega$ of the \textsf{SWIPHT} pulse versus time $t$. The peak amplitude was $\Omega_\textrm{max}/2\pi = \SI{913}{\kHz}$ and the duration was $\tg = \SI{907}{\nano\second}$. \label{fig:pulse-shape}}
\end{figure}

The numerical values in Eqs.~\ref{eq:SWIPHT_length} and \ref{eq:xi_t} are determined by the boundary conditions for the desired $\ket{gg} \leftrightarrow \ket{ge}$ transition  at frequency $\omega_\textnormal{H}$ to complete a $\pi$-rotation and the undesired $\ket{eg} \leftrightarrow \ket{ee}$ transition at frequency $\omega_\textnormal{H} + 2\chi_\textnormal{qq}$ to complete a $2\pi$-rotation. For our experiment, $\omega_\textnormal{d} /2\pi = \SI{6.75427}{\GHz}$, $\Omega_\textrm{max}/2\pi = \SI{913}{\kHz}$, and $\tg = \SI{907}{\nano\second}$ and the resulting control pulse envelope is shown in Fig.~\ref{fig:pulse-shape}. The required \textsf{SWIPHT} pulse was generated using a 25 GSa/s Tektronix\textsuperscript{\textregistered} AWG70002A arbitrary waveform generator (AWG). Since an AWG performs digital signal processing, it has a limited sampling rate and voltage resolution, resulting in minor distortions of the output waveform compared to an ideal waveform. The parameters for the applied pulse were initially calibrated by mapping the effect of \textsf{SWIPHT} on qubit H controlled on the state of qubit L over the 2D landscape of $\tau_\textnormal{g}$ and $\Omega_\textnormal{max}$, and comparing to master equation simulations (see Supplementary material).

\subsection{Joint Qubit Readout \label{ssec:readout}}

\begin{figure}[b]
\centering
\includegraphics[width=\columnwidth]{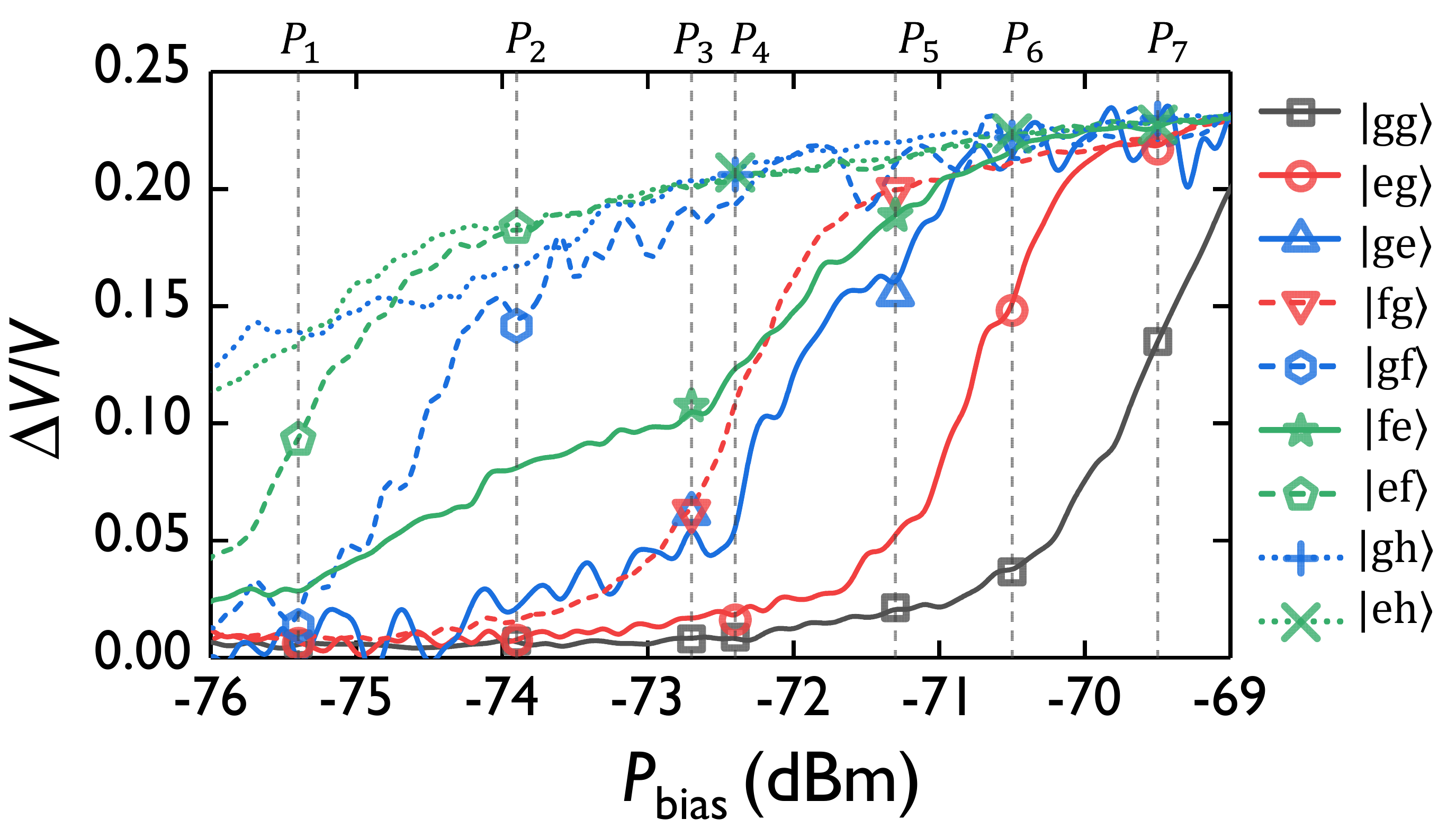}
\caption{Plot of pulsed cavity transmission measurements $\Delta V/V$, averaged over 1000 shots, versus cavity bias power $P_\textrm{bias}$ for different initial two-qubit states. The measurement was performed at the bare cavity frequency $\omega_\textrm{R}/2\pi = \SI{7.680438}{\GHz}$. Dashed lines show powers $P_1$--$P_7$ chosen for joint-qubit readout. Each power $P_i$ has four associated symbols, corresponding to the system being prepared in $\ket{gg}$ (black), $\ket{eg}$ (red), $\ket{ge}$ (blue), or $\ket{ee}$ (green), and then mapped to a higher level in some cases (see Table~\ref{tab:maps}) before being read out. \label{fig:sCurve_biases}}
\end{figure}

Since both of the qubits in our system were coupled to the same 3D cavity, we needed to modify the standard non-linear high-power state-measurement technique \citep{Reed} to perform a joint measurement of the state of both qubits. If only one transmon were coupled to the cavity, a high-power pulse at the bare cavity frequency $\omega_\textrm{R}/2\pi = \SI{7.680438}{\GHz}$ would destructively measure the state of the transmon, with optimization of the pulse power allowing maximum discrimination between the $\ket{g}$ and $\ket{e}$ states. In general, mapping $\ket{e}$ to yet higher states of the transmon will induce transmission through the cavity at a lower power and potentially allow a higher fidelity readout \citep{Premaratne}. This single-qubit method can be extended to perform joint-readout in a two-transmon system by using mappings to higher transmon levels and measuring at multiple carefully chosen powers. The joint-readout was calibrated by first initializing the two-transmon system in each of the states $\ket{gg},~\ket{eg},~\ket{ge},~\ket{fg},~\ket{gf},~\ket{fe},~\ket{ef},~\ket{gh},~\textrm{and}~\ket{eh}$. After each initialization, a $\SI{2}{\us}$ long pulse was applied on resonance with the bare cavity frequency, the transmission through the cavity was measured, and the results were averaged over 1000 shots. Fig.~\ref{fig:sCurve_biases} shows the resulting transmission as a function of the cavity bias power. 

The vertical dashed lines in Fig.~\ref{fig:sCurve_biases} show the seven bias powers $P_1$--$P_7$ we selected for joint qubit-state readout. $\ket{e}\rightarrow \ket{f}$ or $\ket{e}\rightarrow \ket{f}\rightarrow \ket{h}$ mappings were then chosen based on their ability to discriminate between the eigenstates $\ket{gg},~\ket{ge},~\ket{eg},~\textrm{and}~\ket{ee}$. The bias powers were hand-picked to maximize the amount of information extracted using a minimal number of measurements. For example, at power $P_4$, the $\ket{gh}$ and $\ket{eh}$ signals were nearly identical ($\sim 0.21$), and the $\ket{gg}$ and $\ket{eg}$ signals were also close to each other ($\sim 0.01$). Thus, if the excited state of qubit H was mapped $\ket{e}\rightarrow \ket{f} \rightarrow \ket{h}$, $P_4$ would be an ideal bias point to measure the state of transmon H, irrespective of the state of transmon L. In contrast, if qubit H were only mapped $\ket{e}\rightarrow \ket{f}$, the resulting signals for $\ket{gf}$ and $\ket{ef}$ would be further apart. Similarly, if qubit L was mapped $\ket{e}\rightarrow \ket{f}$, the resulting signal is not as large (0.08--0.12) and provides less contrast (see Fig.~\ref{fig:sCurve_biases}). Furthermore, we preferred to use points on Fig.~\ref{fig:sCurve_biases} where the curves had smaller slopes, to reduce the uncertainty in measured values. Table~\ref{tab:maps} summarizes the resulting bias points and mappings that we used.

\begin{figure}[b]
\centering
\includegraphics[width=\columnwidth]{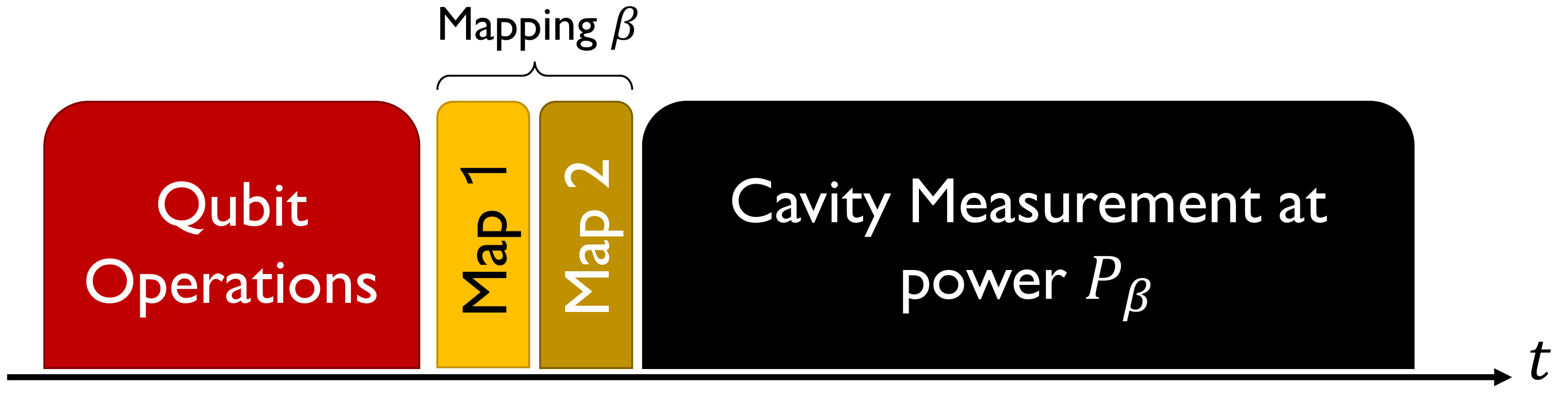}
\caption{Pulse sequence for performing the joint readout (see Table~\ref{tab:maps}) [not to scale]. \label{fig:measure_pulses}}
\end{figure}

\begin{table}[t]
\begin{ruledtabular}
\caption{Parameters for joint-qubit readout shown in Fig.~\ref{fig:measure_pulses}. Seven different cavity input bias powers $P_\beta$ between $\SI{-75.4}{\dBm}$ and $\SI{-69.5}{\dBm}$ were used for joint qubit measurements. Note that in all cases, the $\ket{e}$ state of transmon L or transmon H was mapped to the $\ket{f}$ or $\ket{h}$ state before applying the cavity readout pulse. Also Map 2 was only used for $\beta=4,~6,~\textnormal{and}~7$, to map $\ket{f} \rightarrow \ket{h}$ for $Q_\textnormal{H}$. \label{tab:maps}}
\begin{tabular}{c c c c c}
   \multirow{2}{*}{$\beta$}  & \multirow{2}{*}{$P_\beta$ (dBm)} & \multicolumn{2}{c}{Map 1}  & Map 2 \\
   \cmidrule{3-4} \cmidrule{5-5}
  &  & $Q_\textnormal{L}$ & $Q_\textnormal{H}$ & $Q_\textnormal{H}$ \\
  \midrule
  	1 & $-75.4$ & --- 									& $\textrm{e} \rightarrow \textrm{f}$ 	& --- \\
	2 & $-73.9$ & --- 									& $\textrm{e} \rightarrow \textrm{f}$ 	& --- \\
	3 & $-72.7$ & $\textrm{e} \rightarrow \textrm{f}$ 	& ---  									& --- \\
	4 & $-72.4$ & --- 									& $\textrm{e} \rightarrow \textrm{f}$  	& $\textrm{f} \rightarrow \textrm{h}$\\
	5 & $-71.3$ & $\textrm{e} \rightarrow \textrm{f}$ 	& ---  									& --- \\
	6 & $-70.5$ & --- 									& $\textrm{e} \rightarrow \textrm{f}$  	& $\textrm{f} \rightarrow \textrm{h}$ \\
	7 & $-69.5$ & --- 									& $\textrm{e} \rightarrow \textrm{f}$ 	& $\textrm{f} \rightarrow \textrm{h}$
\end{tabular}
\end{ruledtabular}
\end{table}

Figure~\ref{fig:measure_pulses} shows a schematic our joint measurement procedure: after gating operations on the qubits were finished, upto two state maps were applied, followed by a cavity measurement pulse. For a given initial state, the application of a mapping $\beta$ followed by a measurement pulse at cavity bias power $P_\beta$ results in an average transmission signal $\mathcal{V}_\beta$ (see Appendix~\ref{app:calib} for details). To determine the probability $\mathcal{P}_{nm}$ for an unknown state to be in state $\ketit{nm}$, we first determined the average values of $\overline{\mathcal{V}_\beta}$ for $\beta = \{1, \dots, 7 \}$ by completing $\sim 1000$ repeated preparations, mappings, and measurements of the same state. Then, to find the best fit values for the probabilities $\left\{ \pgg,~\pge,~\peg,~\pee \right\}$, $\chi^2$ minimization was performed on Eq.~\ref{eq:joint_measurement}.


\section{Quantum State Tomography \label{sec:qst}}

\begin{figure}[t]
\centering
\includegraphics[width=0.45\textwidth]{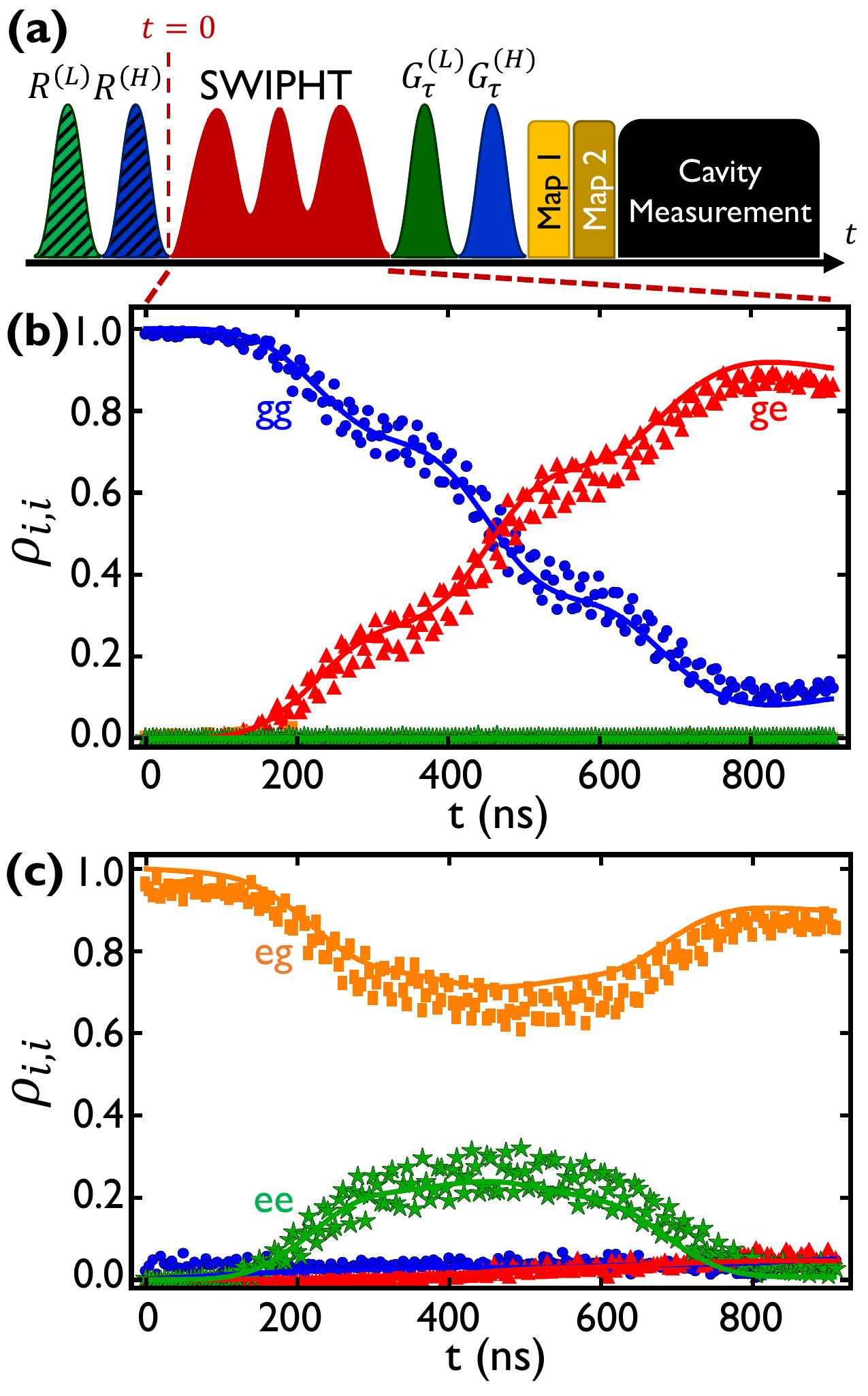}
\caption{(a) Pulse sequence used to perform QST [not to scale]. The two pulses before and after the \textsf{SWIPHT} pulse are used for initializing the qubits and to apply tomographic gates, respectively. (b) Measured populations in $\ket{gg}$ (blue circles), $\ket{ge}$ (red triangles), $\ket{eg}$ (orange rectangles), and $\ket{ee}$ (green stars) from QST, after initializing the system in $\ket{gg}$ and executing the \textsf{SWIPHT} protocol with qubit L being the control qubit. The solid curves are from master equation simulations of the two-qubit system. (c) Measured population during \textsf{SWIPHT} after preparing the system in $\ket{eg}$. Note that time $t=0$ is at the start of the \textsf{SWIPHT} pulse. \label{fig:evolutions}}
\end{figure}

The joint qubit readout described in Sec.~\ref{ssec:readout} provides measured values for $\mathcal{P}_{nm}$, which are the probabilities for the diagonal elements $\rho_{ii}$ of the two-qubit density matrix. The off-diagonal elements of the density matrix were obtained using quantum state tomography (QST), which invokes specific rotations ($G_\tau$) to the state of the system prior to measurement (see Fig.~\ref{fig:evolutions}(a)) \citep{Filipp, Chow2010}. Measurements were obtained using 17 tomographic pulse combinations $G_\tau$ taken repeatedly for each of the seven cavity measurement biases $P_\beta$, resulting in an overcomplete set $\overline{\mathcal{V}_{\beta,\tau}}$ of `measurement' of the system (for details see Appendix~\ref{app:QST_tomoPulses}) \citep{Chow2010}. 

To analyze these results, we used a parameterized explicitly-physical representation of the density matrix $\hat{\rho}$ \citep{James}; this ensures we arrive at a physically consistent density matrix even when state preparation and measurement (SPAM) errors are present. We then use maximum likelihood estimation to obtain the most probable representation of $\hat{\rho}$ \citep{James}, where the likelihood function $\mathcal{L}$ is given by
\begin{align}
\mathcal{L} = \sum_{\tau = 1}^{17} \sum_{\beta= 1}^7 \left[ \frac{\overline{\mathcal{V}_{\beta,\tau}} - \mathcal{M}_{\beta,\tau} \left(G_\tau \hat{\rho} G_\tau^\dagger \right)}{ \sigma_{\overline{\mathcal{V}}, \beta} \left(G_\tau \hat{\rho} G_\tau^\dagger \right) } \right]^2 \, .
\end{align}
Here $\hat{\rho}$ is the parameterized density matrix, $\tau$ is the index for applied tomographic gates $G_\tau$, and $\beta$ is the index for the cavity measurement. The numerator of each term in $\mathcal{L}$ quantifies the closeness of the matrix $\hat{\rho}$ to the data. $\mathcal{M}_{\beta,\tau} (\rho)$ is the analytical expression for the expected average voltage and $\sigma_{\overline{\mathcal{V}}, \beta} (\rho)$ is the estimated uncertainty in measurement of $\overline{\mathcal{V}_{\beta,\tau}}$ (see Appendix~\ref{app:calib} for details). $G_\tau$ are operators that act on $\hat{\rho}$ to generate the tomographic gates used for QST (see Appendix~\ref{app:QST_tomoPulses}). Minimizing $\mathcal{L}$ with respect to the free parameters in $\hat{\rho}$ gives the most probable physical density matrix. 

\subsection{QST during \textsf{SWIPHT} evolution \label{ssec:evolutions}}

To verify that the state of the system was evolving properly during the application of the $\textsf{SWIPHT}$ pulse, we prepared the system in the desired initial state, halted the application of the $\textsf{SWIPHT}$ pulse after a given time $t$, applied tomographic pulses and finally performed a joint-qubit measurement (see Fig.~\ref{fig:evolutions}(a)). At each \SI{5}{\ns} timestep, this process was repeated for 1000 shots for each of the seven cavity biases and each tomographic gate $G_\tau$, to obtain the average responses $\mathcal{M}_{\beta, \tau}$ at time $t$. 

Figure~\ref{fig:evolutions}(b) shows the resulting measured and simulated populations $\rho_{ii}$ versus time $t$, for the \textsf{SWIPHT} protocol, when the system was initialized in the state $\ket{gg}$. We see that upon completion of the \textsf{SWIPHT} pulse, the $\ket{gg}$ state has been placed in the $\ket{ge}$ state with fidelity $\sim 90\%$. The simulations, which include decoherence, are in good agreement with the data, and suggest that the fidelity is being limited by the fact that gate time $\tg = \SI{907}{\ns}$ is a significant fraction of the coherence time of qubit H: $T_{2H} = \SI{6.2}{\us}$. In contrast, Fig.~\ref{fig:evolutions}(c) shows that when the system is initialized in the state $\ket{eg}$, it returns back to $\ket{eg}$ by the end of the \textsf{SWIPHT} pulse, as expected. Note also in Fig.~\ref{fig:evolutions}(c) that the population for $\ket{ee}$ becomes significant but then returns to near zero by the end of the pulse, in good agreement with the simulations.

\subsection{Verification of coherence and phase control}

\begin{figure}[t]
\centering
\includegraphics[width=1\columnwidth]{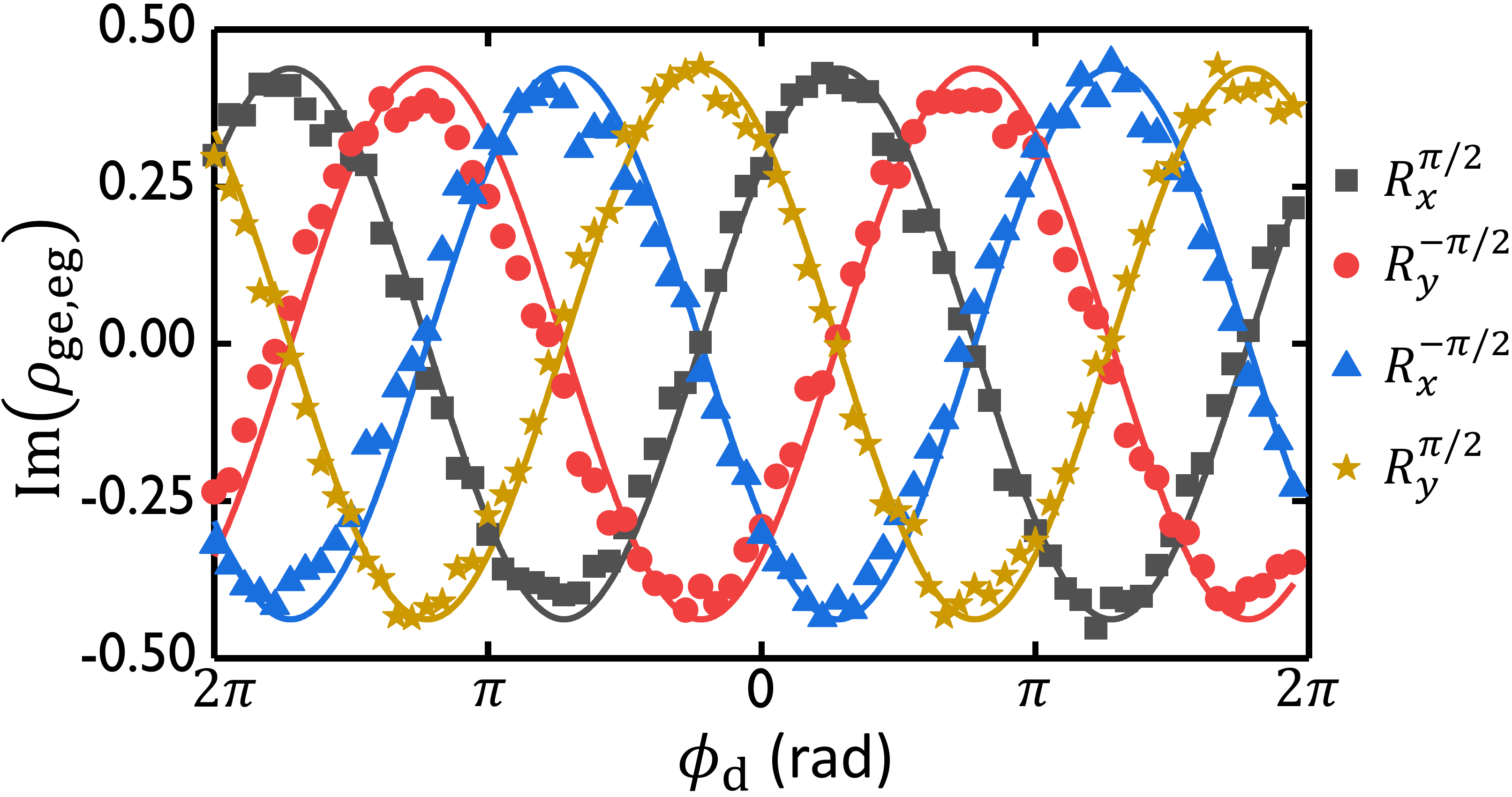}
\caption{Plot of $\textrm{Im} \left(\rho_\textrm{ge,eg} \right)$ versus the phase $\phi_\textrm{d}$ of the \textsf{SWIPHT} pulse, for four different initial superposition states of qubit L $\left(R_{x,y}^{\pm \pi/2} \otimes I \right) \ket{gg}$. Points are from QST measurements and the solid curves are from master equation simulations of $\rho$ with a fixed added phase offset of \SI{245}{\degree}. \label{fig:multiangle}}
\end{figure}

We note that the eigenstate populations shown in Fig.~\ref{fig:evolutions} could have been obtained from our joint qubit readout without performing QST. However, QST also allowed us to construct the entire density matrix and observe the evolution of off-diagonal terms. To verify phase control when using the $\textsf{SWIPHT}$ gate, we varied the drive phase $\phi_\textrm{d}$ from $-2\pi$ to $+2\pi$ and generated different Bell states between the two qubits. The control qubit L was first initialized in one of four superposition states given by $\left(R_{x,y}^{\pm \pi/2} \otimes I \right) \ket{gg}$, where $I$ is the identity operator and $R$ is the qubit rotation operator with a given axis ($x$ or $y$) and phase ($\pm \pi/2$). The $\textsf{SWIPHT}$ gate was then applied and QST performed. Since the initial state had an equal superposition in the $\ket{g}$ and $\ket{e}$ states of the control qubit (L), applying a generalized $\textsf{CNOT}$ gate should generate a Bell state between the two qubits. Figure~\ref{fig:multiangle} shows the variation of the imaginary component of the off-diagonal term $\rho_\textrm{ge,eg}$ versus $\phi_\textrm{d}$. We see a clear oscillatory pattern with period $2\pi$, and an average amplitude of 0.41. We find good agreement with simulations provided we include a phase offset of \SI{245}{\degree}, which is likely due to the length of the input line between the source and cavity.

\section{Quantum process tomography\label{sec:qpt}}

\begin{figure*}[t]
\centering
\includegraphics[width=1\textwidth]{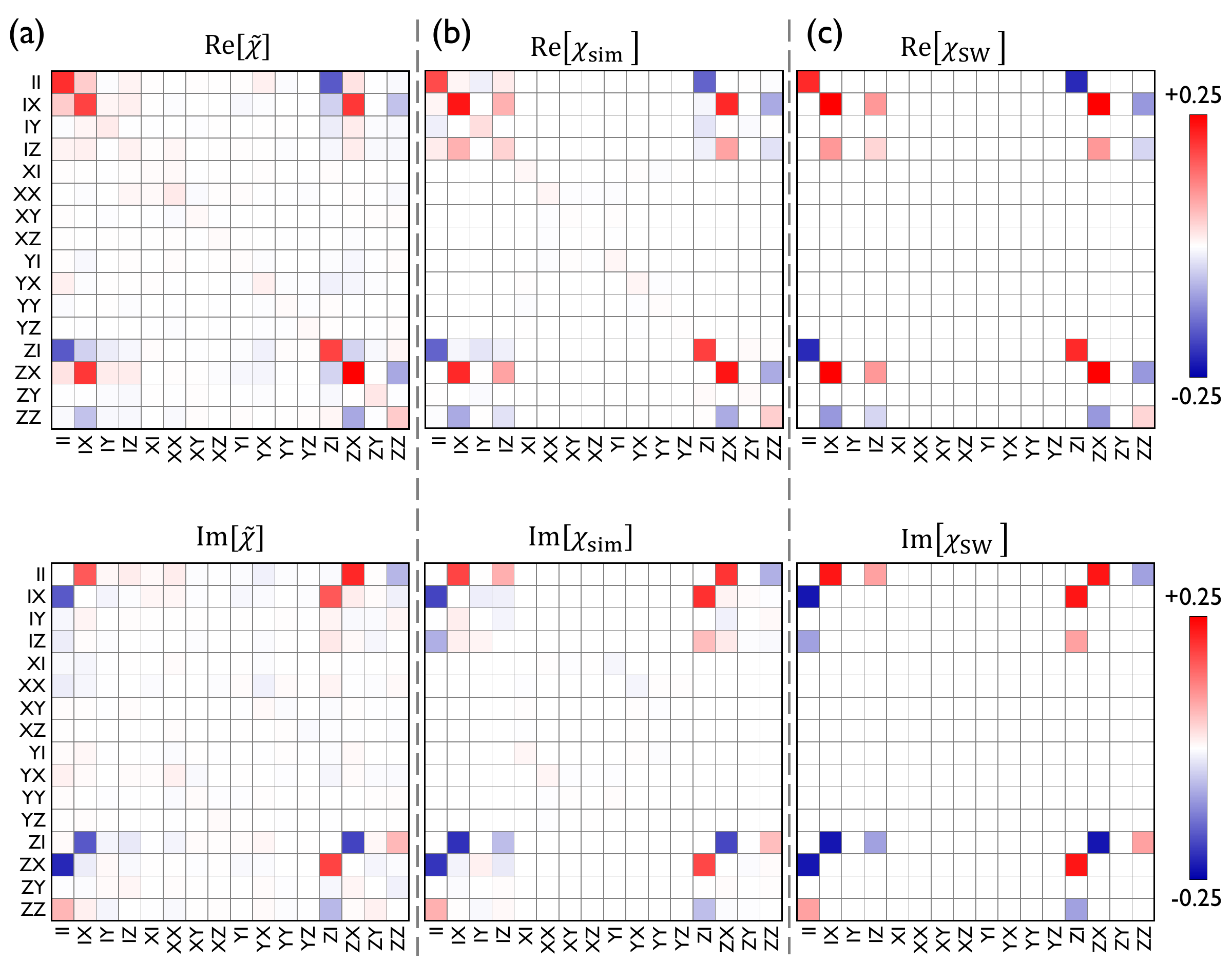}
\caption{(a) Real and imaginary parts of experimental ($\widetilde{\chi}$) , (b) simulated ($\chi_\textnormal{sim}$), and (c) ideal ($\chi_\textsf{SW}$) process matrices $\chi$ for the \textsf{SWIPHT} gate with qubit L as the control. The theoretical maximum magnitude for any element is 0.25. \label{fig:qpt}}
\end{figure*}

To better characterize the \textsf{SWIPHT} gate, quantum process tomography (QPT) was performed \citep{OBrien, Chow2012}. We used QPT to fully characterize our gates, since the relatively long \textsf{CNOT} gate time compared to qubit coherence times rendered randomized benchmarking infeasible \citep{Knill2008,Magesan2011}. The essence of QPT is to prepare a complete set of initial states, perform the desired gate on each state, and then do QST on all the resulting states \citep{Nielsen2014}. We apply the method of \citet{OBrien} and use the $\chi$-matrix representation. The $\chi$-matrix is a positive-definite Hermitian matrix that uniquely describes a quantum process $\hat{\mathcal{O}}$ by mapping any input density matrix to an output density matrix \citep{Nielsen2014}. $\chi$ is implicitly defined via the relation
\begin{align}
\tilde{\rho}^{(\sigma)} = \hat{\mathcal{O}} \left( \rho_0^{(\sigma)} \right) = \sum_{m,n = 1}^{16} \chi_{mn} \hat{A}_m \rho_0^{(\sigma)} \hat{A}_n^\dagger, \label{eq:dm_mapping}
\end{align}
where $\rho_0^{(\sigma)}$ is the input density matrix corresponding to the state $\sigma$, $\tilde{\rho}^{(\sigma)}$ is the output density matrix, and the $\left\{\hat{A}_m \right\}$ are a complete set of quantum gate operators acting on $\rho_0^{(\sigma)}$. For example, for the simple case where the gate operation $\hat{\mathcal{O}}$ is the $j^\textnormal{th}$ basis operation $\hat{A}_j$, one finds $\chi_{jj}=1$ and all other elements of $\chi$ are zero. On the other hand, if $\hat{\mathcal{O}}$ is a non-trivial linear combination of the basis operations, $\chi$ can be thought of as being a look-up table for the weight of each basis operation required to generate $\hat{\mathcal{O}}$. Using the two-qubit Pauli operators $\{\hat{A}_m\} =$ \{$\hat{I}\hat{I}$, $\hat{I}\hat{X}$, $\hat{I}\hat{Y}$, $\hat{I}\hat{Z}$, $\hat{X}\hat{I}$, $\hat{X}\hat{X}$, $\hat{X}\hat{Y}$, $\hat{X}\hat{Z}$, $\hat{Y}\hat{I}$, $\hat{Y}\hat{X}$, $\hat{Y}\hat{Y}$, $\hat{Y}\hat{Z}$, $\hat{Z}\hat{I}$, $\hat{Z}\hat{X}$, $\hat{Z}\hat{Y}$, $\hat{Z}\hat{Z}$\} as a basis for gate operations, an ideal \textsf{CNOT} gate can be written as
\begin{align}
\textsf{CNOT} &= \frac{1}{2}~\hat{I}\hat{I} + \frac{1}{2}~\hat{I}\hat{X} - \frac{1}{2}~\hat{Z}\hat{I} + \frac{1}{2}~\hat{Z}\hat{X}.
\end{align}
Similarly the \textsf{SWIPHT} operation is 
\begin{align}
\textsf{SWIPHT} &= i0.4577~\hat{I}\hat{I} + \frac{1}{2}~\hat{I}\hat{X} -i0.4577~\hat{Z}\hat{I} + \frac{1}{2}~\hat{Z}\hat{X} \nonumber \\ 
 &  + 0.2012~\hat{I}\hat{Z} - 0.2012~\hat{Z}\hat{Z},
\end{align}
where contributions to the norm smaller than $10^{-4}$ have been neglected in the real components of the coefficients for $\hat{I}\hat{I}$ and $\hat{Z}\hat{I}$, and the imaginary components of the coefficients for $\hat{I}\hat{Z}$ and $\hat{Z}\hat{Z}$. The resulting $\chi$-matrix has 36 non-zero elements, 20 of which are real and 16 which are imaginary (see Fig.~\ref{fig:qpt}(c)). The visual representation of the $\chi$-matrix can serve as a gate diagnostic tool, with discrepancies between ideal and experimental $\chi$-matrices providing insight into sources of systematic errors and possible remedies \citep{Nielsen2014}.

We use QPT to extract the $\chi$-matrix from our measurements. As in QST, for our QPT analysis, we explicitly restrict the form of the $16\times 16$ $\chi$-matrix so that it is physical \citep{OBrien}. To ensure that $\chi$ represents a trace-preserving process, \textit{i.e.} the density matrix remains Hermitian with trace one, the completeness relation $\sum_{mn} \chi_{mn} \hat{A}_n^\dagger \hat{A}_m = \hat{I}$ must be satisfied \citep{Nielsen2014, OBrien}. The completeness relation can be enforced by requiring that
\begin{align}
\Xi = \sum_{i,j = 1}^{16} \left\{ \left[\textrm{Re} (\mathcal{Z}_{ij}) - I_{ij} \right]^2  + \left[\textrm{Im} (\mathcal{Z}_{ij}) \right]^2 \right\}
\end{align}
is minimized as a function of each complex $\chi_{ij}$, where we have defined $\hat{\mathcal{Z}} = \sum_{mn} \chi_{mn} \hat{A}_n^\dagger \hat{A}_m$. To find the best fit $\chi_{ij}$ subject to the completeness constraint, we introduce a Lagrange multiplier $\lambda$ and minimize the likelihood function
\begin{align}
\widetilde{\mathcal{L}} = \sum_{\sigma = 1}^{36} \sum_{\tau = 1}^{17} \sum_{\beta= 1}^7 \left[ \frac{\overline{\mathcal{V}_{\beta,\tau}^{(\sigma)}} - \mathcal{M}_{\beta,\tau} \left(G_\tau \tilde{\rho}^{(\sigma)} G_\tau^\dagger \right)}{ \sigma_{\overline{\mathcal{V}}, \beta} \left(G_\tau \tilde{\rho}^{(\sigma)} G_\tau^\dagger \right) } \right]^2 + \lambda \Xi \, , \label{eq:qptEquation}
\end{align}
with respect to $\lambda$ and all $\chi_{mn}$. Here $\sigma$ is an index for the overcomplete set of 36 initial states given by $\{ I, R_x^{\pi}, R_{x,y}^{\pm \pi/2} \}^{\otimes 2} \ket{gg}$, $\tilde{\rho}^{(\sigma)}$ is given by Eq.~\ref{eq:dm_mapping}, and each $\overline{\mathcal{V}_{\beta,\tau}^{(\sigma)}}$ is obtained from the average of 1000 measurements. 

Minimizing $\widetilde{\mathcal{L}}$ with respect to each complex quantity $\chi_{mn}$ and $\lambda$ yields $\widetilde{\chi}$, the maximum-likelihood estimate for the $\chi$-matrix that describes the underlying process. Fig.~\ref{fig:qpt}(a) shows $\widetilde{\chi}$ we extracted from our measurements of the \textsf{SWIPHT} gate. Fig.~\ref{fig:qpt}(b) and Fig.~\ref{fig:qpt}(c) show respectively, the $\chi$-matrices obtained for master equation simulations of the \textsf{SWIPHT} gate incorporating decoherence ($\chi_\textnormal{sim}$) and the ideal decoherence-free case ($\chi_\textsf{SW}$). The effects of decoherence can be seen by comparing $\chi_\textnormal{sim}$ to the ideal $\chi_\textsf{SW}$; for example all the non-zero components in $\chi_\textsf{SW}$ are slightly smaller in magnitude in $\chi_\textnormal{sim}$. The presence of some additional non-zero components in $\widetilde{\chi}$ and $\chi_\textnormal{sim}$, that do not occur in $\chi_\textsf{SW}$, indicate the presence of coherent errors \citep{Kaye2007}. For example, the real part of $\widetilde{\chi}$ and $\chi_\textnormal{sim}$ have non-zero contributions from $\left( II, IX \right)$ and $\left( IX, II \right)$, while the contributions vanish in the ideal $\chi_\textnormal{SW}$. In addition, by comparing $\widetilde{\chi}$ with $\chi_\textnormal{sim}$ we can infer that additional coherent $z$-rotations are present in $\widetilde{\chi}$. We note that we also observed extra $z$-rotations using QPT on single qubit gates (see Supplementary material).

To compare our extracted $\widetilde{\chi}$ to the expected $\chi_\textsf{SW}$ from the ideal decoherence-free \textsf{SWIPHT} operation, we compute the process fidelity
\begin{align}
\mathcal{F}_\textrm{p} &= \textrm{Tr}\left(\widetilde{\chi} \chi_\textsf{SW}  \right), \label{eq:processFidelity}
\end{align}
and the mean gate fidelity
\begin{align}
\mathcal{F}_\textrm{g} &= \frac{\mathcal{F}_\textrm{p} d +1}{d+1}, \label{eq:gateFidelity}
\end{align}
where $d=4$ is the dimensionality of our two-qubit system \citep{OBrien, Gilchrist, Nielsen2002}. We find that $\mathcal{F}_\textrm{p} = 0.79$, $\mathcal{F}_\textrm{g} = 0.83$, and the average purity of the gate $\overline{\textrm{Tr} \left( \rho^2 \right)} = \left[ d \textrm{ Tr}\left( \widetilde{\chi}^2 \right) + 1 \right] / (d+1) = 0.75$ \citep{OBrien}. Average purity is a metric for the degree of mixture introduced by a gate and indicates how much the process was affected by incoherent errors \citep{Gilchrist}.

\begin{table}[b]
	\begin{ruledtabular}
	\caption{Gate times $\tau_\textnormal{gate}$ and measured QPT performance metrics $\mathcal{F}_\textrm{p}$, $\mathcal{F}_\textrm{g}$, and $\overline{\textrm{Tr} \left( \rho^2 \right)}$ for several gates. For the four \textsf{SWIPHT} gates at the bottom of the table, the subscript denotes the control qubit, and the superscript denotes the phase $\phi_\textnormal{d}$ of the drive. \label{tab:qpt_performance}}
		\begin{tabular}{l c c c c}
		  	Gate & $\tau_\textnormal{gate}$ (ns) & $\mathcal{F}_\textrm{p}$ & $\mathcal{F}_\textrm{g}$ &  $\overline{\textrm{Tr} \left( \rho^2 \right)}$ \\
		  	\midrule
		  	$I \otimes I$ 								& 0 	& 0.97 	& 0.97 	& $>0.99$ \\
		  	$I \otimes R_x^{\pi}$ 						& 37 	& 0.95 	& 0.96 	& $>0.99$ \\
		  	$I \otimes R_y^{\pi}$ 						& 37 	& 0.94 	& 0.95 	& 0.99 \\
		  	$I \otimes R_x^{\pi/2}$ 					& 37 	& 0.94 	& 0.96 	& $>0.99$ \\
		  	$I \otimes R_y^{\pi/2}$ 					& 37 	& 0.92	& 0.93 	& 0.99 \\
		  	$I \otimes R_x^{-\pi/2}$ 					& 37 	& 0.93 	& 0.95 	& $>0.99$ \\		  	
		  	$I \otimes R_y^{-\pi/2}$ 					& 37 	& 0.93 	& 0.94 	& 0.99\\
		  	$R_x^{\pi} \otimes I$ 						& 72 	& 0.92 	& 0.93 	& 0.99 \\
		  	$R_y^{\pi} \otimes I$ 						& 72 	& 0.90 	& 0.92 	& 0.98 \\
		  	$R_x^{\pi/2} \otimes I$ 					& 72 	& 0.89 	& 0.91 	& 0.98 \\
		  	$R_y^{\pi/2} \otimes I$ 					& 72 	& 0.87 	& 0.90 	& 0.98 \\
		  	$R_x^{-\pi/2} \otimes I$ 					& 72 	& 0.84 	& 0.87 	& 0.98 \\
		  	$R_y^{-\pi/2} \otimes I$					& 72 	& 0.87 	& 0.90 	& $>0.99$ \\
		  	$\textsf{SWIPHT}^0_{\textrm{L}}$ 			& 907 	& 0.79 	& 0.83 	& 0.75\\
		  	$\textsf{SWIPHT}^0_{\textrm{H}}$ 			& 907 	& 0.79	& 0.83 	& 0.77 \\
		  	$\textsf{SWIPHT}^{49\pi/36}_{\textrm{L}}$ 	& 907 	& 0.80 	& 0.84 	& 0.75\\
		  	$\textsf{SWIPHT}^{7\pi/12}_{\textrm{H}}$ 	& 907 	& 0.80 	& 0.84 	& 0.77
		\end{tabular}
	\end{ruledtabular}
\end{table}

To better understand how the fidelity of our \textsf{SWIPHT} gate was limited by decoherence, we also examined the fidelity obtained from $\chi_\textnormal{sim}$. $\chi_\textnormal{sim}$ was found by first performing density matrix simulations of the \textsf{SWIPHT} operation in the presence of decoherence, and then performing QPT on the simulated results (see Fig.~\ref{fig:qpt}(b)). The master equation simulations were run with a complete set of 16 initial states and the resulting final density matrices were used to construct the likelihood function for QPT. To simplify the minimization procedure, we assumed an equal numerical error for all simulated ``measurements''. From $\chi_\textnormal{sim}$, we found that the expected values for the performance metrics in the presence of decoherence were $\mathcal{F}_\textrm{p} = 0.84$, $\mathcal{F}_\textrm{g} = 0.87$, and $\overline{\textrm{Tr} \left( \rho^2 \right)} = 0.77$.  Comparing with the experiment, we see that most of the infidelity was due to the gate time $\tg$ not being sufficiently short compared to the $T_2$ of the devices, with the simulations giving fidelities 3--4\% greater than data. We also performed master equation simulations on an identical system with qubit lifetimes set to $T_1 = \SI{40}{\us}$; the gate fidelity was found to be $\mathcal{F}_\textrm{g} = 0.98$, further indicating that at present the primary limiting factor is the length of the gate time compared to the coherence times.

To understand why the \textsf{SWIPHT} gate gave somewhat lower fidelity than found in the simulation, we also performed QPT on other gates, including three simple variations on the \textsf{SWIPHT} gate (see Table~\ref{tab:qpt_performance} and Supplementary material). All our single qubit gates had high purity due to fast gate times, while all the \textsf{SWIPHT} gates showed lower purity, in part due to qubit relaxation during the longer gate times.  Of particular note was the $I \otimes I$ gate, which was set up to be a zero-length operation undergoing QPT. For this simple null gate we found $\mathcal{F}_\textnormal{g}=0.97$, supporting that there were $\sim 3\%$ errors introduced due to SPAM during QPT. Note that the $I \otimes I$ operation had the highest performing metrics compared to all other gates (see Table~\ref{tab:qpt_performance}).

The four \textsf{SWIPHT} gates included two pairs of gates utilizing either qubit L or qubit H as the control. Within each pair, different initial drive phases were employed to verify the degree of phase control on the gate. Given the similarities in fidelity between all four variations, this indicated no preferred direction and good phase control for the \textsf{CNOT} gate.

The single qubit gates were executed by applying single-tone Gaussian pulses either at the $\ket{gg}\leftrightarrow \ket{eg}$ frequency or the $\ket{gg}\leftrightarrow \ket{ge}$ frequency for qubits L and H, respectively. Due to the relatively large $\chi_\textnormal{qq}$ shift, the single-qubit transitions $\ket{ge}\leftrightarrow \ket{ee}$ and $\ket{eg}\leftrightarrow \ket{ee}$ were inevitably driven off-resonantly. Although the effective drive strength was high, the frequency shift inevitably resulted in control errors and relatively low gate fidelities for some single qubit gates (see Table~\ref{tab:qpt_performance}).


\section{Conclusions \label{sec:conclusions}}

In this experiment, we demonstrated a generalized $\textsf{CNOT}$ operation between two fixed-frequency transmon qubits using the \textsf{SWIPHT} technique \citep{Economou, Deng}. Four slightly different implementations of \textsf{SWIPHT} yielded gate fidelities $\mathcal{F}_\textrm{g} \simeq 83-84\%$, similar to initial demonstrations of other all-microwave gates \citep{Chow2013,Chow2011,Poletto2012}. These fidelities were 3--4\% less than fidelities obtained from master equation simulations that incorporated decoherence, most likely due to SPAM errors as discussed in Sec.~\ref{sec:qpt}. We note that increased sensitivity resulting from higher qubit coherence times would have enabled identification of other possible error channels. The presence of the cavity, with a lifetime $1/\kappa = \SI{50}{ns}$, directly affected qubit lifetimes due to the Purcell effect, but otherwise had a minimal impact on the fidelity of the gate.

As proposed by \citet{Deng}, the gate time could be significantly reduced by using qubits with frequencies that are closer together, leading to significant improvements in fidelity. As we noted above, our measured fidelities include errors associated with state preparation and measurement. This suggests that our gate fidelities could have been improved by using pulse-shaping techniques or optimal control techniques during both the gate and QPT \citep{Motzoi, Glaser2015} . Measurements of the two-qubit states were performed using a joint-readout technique which could be improved or replaced by a more efficient technique \citep{Paik2016}. We also found that we could achieve excellent control of the phase of the target qubit by adjusting the phase of the drive. Finally, we note that the residual phase accumulation on the control qubit which we observed during \textsf{SWIPHT} can be corrected through single-qubit rotations to implement a conventional \textsf{CNOT} gate.

\acknowledgments

We thank S. Economou, E. Barnes and X.-H. Deng for helpful comments on the \textsf{SWIPHT} protocol. F.C.W. acknowledges support from the Joint Quantum Institute and the Center for Nanophysics and Advanced Materials. 


\appendix

\section{Details of the cQED system and the system Hamiltonian \label{app:hamiltonian}}

Our cQED system consisted of two transmons embedded in a 3D cavity that was cooled on the mixing chamber of a Leiden CF-450 dilution refrigerator with a base temperature of 20 mK (see Supplementary material for the full details). Each transmon \citep{Paik2011} had a single Al/AlO$_{\textrm{x}}$/Al Josephson junction and two large $\SI{500}{\micro\meter} \times \SI{650}{\micro\meter}$ Al pads capacitively shunting the junction. The Hamiltonian for the coupled system can be written as,
\begin{align}
\mathcal{H}_0 &= \hbar \omega_\textrm{R} a_\textrm{R}^\dagger a_\textrm{R} + \hbar  \sum_{j=\textnormal{L, H}} \left[ \omega_j a_j^\dagger a_j  + \frac{E_\textrm{C}^{(j)}}{2\hbar} a_j^\dagger a_j \left(a_j^\dagger a_j - 1 \right) \right. \nonumber \\ 
&\hspace{1em} \left. + g_j \left( a_\textrm{R}^\dagger a_j + a_\textrm{R} a_j^\dagger \right)  \right] + \hbar J \left( a_\textrm{L}^\dagger a_\textrm{H} + a_\textrm{L} a_\textrm{H}^\dagger \right), \label{eq:full_H}
\end{align}
where $a$ and $a^\dagger$ represent the annihilation and creation operators and the other relevant system parameters are listed in Table~\ref{tab:system_parameters}.

\begin{table}[h]
	\begin{ruledtabular}
	\caption{Device parameters for the two qubits coupled to the measurement cavity. Values were obtained by extracting eigenstates for the Hamiltonian Eq.~\ref{eq:full_H} and comparing the resulting transition frequencies to the measured spectrum. \label{tab:system_parameters}}
		\begin{tabular}{l c c}
		  	System parameter & Symbol & Value \\
		  	\midrule
		  	Bare frequency of transmon L 			& $\omega_L/2\pi$ & \SI{6.10322}{\giga\hertz} \\
		  	Bare frequency of transmon H 			& $\omega_H/2\pi$ & \SI{6.79943}{\giga\hertz} \\
		  	Bare frequency of cavity 				& $\omega_R/2\pi$ & \SI{7.66927}{\giga\hertz} \\
		  	Charging energy of transmon L 			& $E_\textrm{C}^\textrm{(L)}/h$ & \SI{206.5}{\mega\hertz} \\
		  	Charging energy of transmon H 			& $E_\textrm{C}^\textrm{(H)}/h$ & \SI{192.6}{\mega\hertz} \\
		  	Transmon-transmon direct coupling 		& $J/2\pi$ & \SI{14.3}{\mega\hertz} \\
		  	Cavity-transmon coupling (L) 			& $g_\textrm{L}/2\pi$ & \SI{224.6}{\mega\hertz} \\
		  	Cavity-transmon coupling (H) 			& $g_\textrm{H}/2\pi$ & \SI{207.5}{\mega\hertz} \\
		  	Cavity-transmon dispersive shift (L)	& $\chi_\textrm{L}/2\pi$ & \SI{-3.3}{\mega\hertz} \\
		  	Cavity-transmon dispersive shift (H)	& $\chi_\textrm{H}/2\pi$ & \SI{-7.2}{\mega\hertz} \\
		  	Qubit-qubit dispersive shift			& $\chi_\textnormal{qq}/2\pi$ & \SI{-1.04}{\MHz} \\
		  	
		  	Cavity lifetime 						& $1/\kappa$ & \SI{50}{\nano\second} \\
		  	Relaxation time of transmon L 			& $T_\textrm{1L}$ & $\SI{9.0}{\micro\second}$ \\
		  	Relaxation time of transmon H 			& $T_\textrm{1H}$ & $\SI{3.5}{\micro\second}$ \\
		  	Spin-echo time of transmon L 			& $T_\textrm{2L}$ & $\SI{14.6}{\micro\second}$ \\
		  	Spin-echo time of transmon H 			& $T_\textrm{2H}$ & $\SI{6.2}{\micro\second}$ \\
		\end{tabular}
	\end{ruledtabular}
\end{table}

\section{Joint qubit readout calibration \label{app:calib}}

We consider the joint state readout of a two-qubit system with computational basis eigenstates $\left\{ \ket{gg}, \ket{eg}, \ket{ge}, \ket{ee} \right\}$, which we will denote simply as $\nu=1,2,3,4$. We are using a high-power readout and have chosen 7 readout mappings (combinations of bias power and qubit mappings) which we will label with $\beta=1,2,...,7$. If the system is prepared in the state $\nu$ and measured at power $P_\beta$, we will assume that the amplitude $\mathcal{V} \equiv V/\Delta V$ of the transmission is a random variable that is drawn from a probability distribution given by:

\begin{widetext}
\begin{align}
\dv{p^{(\beta)}_\nu}{\mathcal{V}} =K \left\{ \frac{\left(1 - \mathcal{K}_\nu^{(\beta)}\right)}{\sigma_\textnormal{low}^{(\beta, \nu)} \sqrt{2\pi}} \exp \left[-\frac{\left(\mathcal{V}-V_\textnormal{low}^{(\beta, \nu)}\right)^2}{2 \left(\sigma_\textnormal{low}^{(\beta, \nu)}\right)^2} \right] + \frac{\left(\mathcal{K}_\nu^{(\beta)}\right)}{{\sigma_\textnormal{high}^{(\beta, \nu)} \sqrt{2\pi}}} \exp \left[-\frac{\left(\mathcal{V}-V_\textnormal{high}^{(\beta, \nu)}\right)^2}{2 \left(\sigma_\textnormal{high}^{(\beta, \nu)}\right)^2} \right] \right\}, \label{eq:prob_distr}
\end{align}
\end{widetext}
where  $\left(\displaystyle\dv{p^{(\beta)}_\nu}{\mathcal{V}} \right) \dd{\mathcal{V}}$ is the probability of a single-shot measurement producing an output between $\mathcal{V}$ and $\mathcal{V} + \delta \mathcal{V}$, where $\mathcal{V} \equiv \Delta V/V$, and $K$ is a normalizing constant; here $K=1$ since $\int \left(\dv{p}{\mathcal{V}} \right) \dd{\mathcal{V}} =1$. This model takes into account the noise in the readout electronics and the bifurcating nature of the non-linear readout \citep{Reed}. $\mathcal{K}_\nu^{(\beta)}$ and $\left(1 - \mathcal{K}_\nu^{(\beta)} \right)$ are probabilities that the cavity bifurcates to the ``high'' and ``low'' transmission states when measured, respectively. $V_\textnormal{low}^{(\beta, \nu)}$ and $\sigma_\textnormal{low}^{(\beta, \nu)}$ are the mean and standard deviation of the cavity signal $\Delta V/V$ when it bifurcates to the ``low'' transmission state, respectively. $V_\textnormal{high}^{(\beta, \nu)}$ and $\sigma_\textnormal{high}^{(\beta, \nu)}$ are the mean and standard deviation of the cavity signal $\Delta V/V$ when it bifurcates to the ``high'' transmission state, respectively. Note that the underlying distributions corresponding to ``high'' and ``low'' transmission states are assumed to be normal distributions as one would expect for simple added noise.

In general, the system may be prepared in a superposition of states, with probabilities $\mathcal{P}_\nu \equiv \rho_{\nu \nu}$ to be measured in each state $\nu$, where $\rho$ is the density matrix in the computational subspace. Reading out the state of the system with mapping $\beta$ will then produce an output signal $\mathcal{V}$ that is randomly drawn from a distribution that is given by:
\begin{align}
\dv{p^{(\beta)}}{\mathcal{V}} = \sum_{\nu=1}^4 \rho_{\nu \nu} \dv{p^{(\beta, \nu)}_\nu}{\mathcal{V}},
\end{align}
where $\rho_{\nu \nu}$ is the $\nu^\textnormal{th}$ diagonal element of the density matrix in the computational basis. 

\begin{widetext}
If the state is prepared and measured $N$ times using mapping $\beta$, then for large $N$, the average of the resulting $N$ output voltages will converge to:
\begin{align}
\overline{\mathcal{V}_\beta} = \mathcal{M}_\beta (\rho) = \int_{-\infty}^\infty V \dv{p^{(\beta)}}{\mathcal{V}} \dd{\mathcal{V}} = \sum_{\nu=1}^4 \rho_{\nu \nu} \int_{-\infty}^\infty V \dv{p^{(\beta, \nu)}}{\mathcal{V}} \dd{\mathcal{V}} = \sum_{\nu=1}^4 \rho_{\nu \nu} \left[ \left(1 - \mathcal{K}_\nu^{(\beta)}\right) V_\textnormal{low}^{(\beta, \nu)} + \mathcal{K}_\nu^{(\beta)} V_\textnormal{high}^{(\beta, \nu)} \right]. \label{eq:joint_measurement}
\end{align}

With average voltage $\overline{\mathcal{V}_\beta}$ found from the average of $N$ measurements of the signal $\mathcal{V}$, the uncertainty in $\mathcal{V}$ for mapping $\beta$ will then be:
\begin{align}
\sigma_{\overline{\mathcal{V}}, \beta} = \sigma_{\overline{\mathcal{V}}, \beta} (\rho) = \frac{\sigma_{\mathcal{V}, \beta}}{\sqrt{N}} = \sqrt{\sum_{\nu=1}^4 \frac{\rho_{\nu \nu}}{N}  \left\{ \left(1 - \mathcal{K}_\nu^{(\beta)}\right) \left[ \left(\sigma_\textnormal{low}^{(\beta, \nu)}\right)^2 + \left(V_\textnormal{low}^{(\beta, \nu)} - \overline{\mathcal{V}} \right)^2 \right] + \mathcal{K}_\nu^{(\beta)} \left[ \left(\sigma_\textnormal{high}^{(\beta, \nu)}\right)^2 + \left(V_\textnormal{high}^{(\beta, \nu)} - \overline{\mathcal{V}} \right)^2 \right] \right\}}.
\end{align}
\end{widetext}

According to the measurement model described above, the calibration matrices $\mathcal{K}$, $V_\textnormal{low}$, $V_\textnormal{high}$, $\sigma_\textnormal{low}$, and $\sigma_\textnormal{high}$ for the seven cavity bias powers $P_\beta$ and each of the initialized eigenstates $\nu$ were measured to be as follows:
\begin{align}
\mathcal{K} = 10^{-2} \left(
\begin{array}{SSSS}
 0.121 & 7.85 & 0.529 & 43.8 \\ 
 0.509 & 57.0 & 1.40 & 85.8 \\ 
 1.22 & 22.0 & 45.4 & 91.9 \\ 
 2.09 & 90.5 & 8.84 & 94.3 \\ 
 8.28 & 74.5 & 89.3 & 98.6 \\ 
 14.9 & 97.2 & 73.4 & 99.5 \\ 
 61.0 & 99.0 & 97.3 & 99.9
\end{array}
\right), \label{eq:param1}
\end{align}

\begin{align}
V_\textnormal{low} = 10^{-3} \left(
\begin{array}{SSSS}
 4.74 & 5.58 & 4.91 & 5.11 \\ 
 4.38 & 5.48 & 4.88 & 5.50 \\ 
 4.39 & 5.16 & 6.29 & 4.82 \\ 
 3.91 & 4.67 & 4.67 & 4.67 \\ 
 4.22 & 5.56 & 5.16 & 4.22 \\ 
 4.05 & 4.05 & 6.04 & 4.05 \\ 
 4.50 & 4.50 & 4.50 & 4.50
\end{array}
\right),
\end{align}

\begin{align}
V_\textnormal{high} = 10^{-3} \left(
\begin{array}{rrrr}
 213 & 213 & 213 & 213 \\ 
 219 & 219 & 219 & 221 \\ 
 221 & 221 & 218 & 226 \\ 
 227 & 227 & 227 & 227 \\ 
 226 & 228 & 229 & 231 \\ 
 227 & 233 & 229 & 233 \\ 
 231 & 234 & 233 & 235
\end{array}
\right),
\end{align}

\begin{align}
\sigma_\textnormal{low} = 10^{-2} \left(
\begin{array}{SSSS}
 3.37 & 3.39 & 3.47 & 3.99 \\ 
 2.85 & 2.83 & 2.90 & 3.36 \\ 
 2.47 & 2.47 & 2.90 & 2.80 \\ 
 2.79 & 3.22 & 4.65 & 6.45 \\ 
 2.11 & 2.14 & 2.29 & 2.11 \\ 
 2.08 & 2.08 & 2.66 & 2.08 \\ 
 1.89 & 1.89 & 1.89 & 1.89
\end{array}
\right),
\end{align}

\begin{align}
\sigma_\textnormal{high} = 10^{-2} \left(
\begin{array}{SSSS}
 4.05 & 4.05 & 4.05 & 4.05 \\ 
 2.96 & 2.96 & 2.96 & 3.30 \\ 
 2.48 & 2.48 & 3.13 & 2.87 \\ 
 3.13 & 3.13 & 6.56 & 5.49 \\ 
 2.21 & 2.19 & 2.32 & 2.55 \\ 
 2.17 & 2.44 & 2.71 & 4.44 \\ 
 1.92 & 2.13 & 2.20 & 3.98
\end{array}
\right). \label{eq:param5}
\end{align}

Thus for example, $\mathcal{K}_{3}^{(5)} = 0.893$ is the probability for the cavity signal to bifurcate to the ``high'' state when the system is in state $\ket{eg}$ corresponding to $\nu=3$, Map 1 is applied to take $Q_\textnormal{L}$ from $\ket{e} \rightarrow \ket{f}$, and the measurement pulse power is $P_5 = \SI{-71.3}{\dBm}$ corresponding to $\beta=5$. 

The above results were obtained by performing \num{50000} single shot measurements at each of the mappings $\beta$ for each eigenstate $\nu$. Histograms of the resulting single shots were generated and subsequently fitted using Eq.~\ref{eq:prob_distr} to determine the parameters in Eq.~\ref{eq:param1}--\ref{eq:param5}. For certain eigenstates, when the bifurcation to the ``low'' or ``high'' state was overwhelming, fitting was difficult for the diminished component. In such situations, appropriate values from fitting a more balanced histogram at the same mapping $\beta$ were used during the fitting procedure.

\section{Tomographic pulses for QST \label{app:QST_tomoPulses}}

Table~\ref{tab:QSTMaps} gives the 17 tomographic maps we used for QST.

\begin{table}[b]
		\caption{Gate names and corresponding operations. \label{tab:QSTMaps}}
	\begin{minipage}[t]{.20\textwidth}
	\centering
		\begin{ruledtabular}
			\begin{tabular}{c c}
			  	Gate & Tomographic\\
			  	name & map \\
				\cmidrule{1-2}
			  	$G_1$ & $I \otimes I$ \\
			  	$G_2$ & $I \otimes R_x^{\pi/2}$ \\
			  	$G_3$ & $I \otimes R_y^{\pi/2}$ \\
			  	$G_4$ & $R_x^{\pi/2} \otimes I$ \\
			  	$G_5$ & $R_y^{\pi/2} \otimes I$ \\
			  	$G_6$ & $R_x^{\pi/2} \otimes R_x^{\pi/2}$ \\
			  	$G_7$ & $R_x^{\pi/2} \otimes R_y^{\pi/2}$ \\
			  	$G_8$ & $R_y^{\pi/2} \otimes R_x^{\pi/2}$ \\
			  	$G_9$ & $R_y^{\pi/2} \otimes R_y^{\pi/2}$ 
			\end{tabular}
		\end{ruledtabular}
	\end{minipage}
	\hspace{0.05\textwidth}
	\begin{minipage}[t]{.20\textwidth}
	\centering
		\begin{ruledtabular}
			\begin{tabular}{c c}
			  	Gate & Tomographic\\
			  	name & map \\
				\cmidrule{1-2}
			  	$G_{10}$ & $I \otimes R_x^{-\pi/2}$ \\
			  	$G_{11}$ & $I \otimes R_y^{-\pi/2}$ \\
			  	$G_{12}$ & $R_x^{-\pi/2} \otimes I$ \\
			  	$G_{13}$ & $R_y^{-\pi/2} \otimes I$ \\
			  	$G_{14}$ & $R_x^{-\pi/2} \otimes R_x^{-\pi/2}$ \\
			  	$G_{15}$ & $R_x^{-\pi/2} \otimes R_y^{-\pi/2}$ \\
			  	$G_{16}$ & $R_y^{-\pi/2} \otimes R_x^{-\pi/2}$ \\
			  	$G_{17}$ & $R_y^{-\pi/2} \otimes R_y^{-\pi/2}$ \\
			  	\\
			\end{tabular}
		\end{ruledtabular}
	\end{minipage}
\end{table}

\end{document}


\title{Supplemental Material : Implementation of a generalized \textsf{CNOT} gate between fixed-frequency transmons}

\author{Shavindra P. Premaratne}
\email{shavi@umd.edu}
\affiliation{Department of Physics, University of Maryland, College Park, Maryland 20742, USA}
\affiliation{Laboratory for Physical Sciences, College Park, Maryland 20740, USA}
\author{Jen-Hao Yeh}
\affiliation{Department of Physics, University of Maryland, College Park, Maryland 20742, USA}
\affiliation{Laboratory for Physical Sciences, College Park, Maryland 20740, USA}
\author{F. C. Wellstood}
\affiliation{Department of Physics, University of Maryland, College Park, Maryland 20742, USA}
\affiliation{Joint Quantum Institute and Center for Nanophysics and Advanced Materials, College Park, Maryland 20742, USA}
\author{B. S. Palmer}
\affiliation{Laboratory for Physical Sciences, College Park, Maryland 20740, USA}

\date{\today}

\maketitle

\vfill
\hfill
\clearpage
\newpage

\section{Schematic of the complete control and measurement network}

\begin{figure}[h]
\centering
\includegraphics[width=0.76\columnwidth]{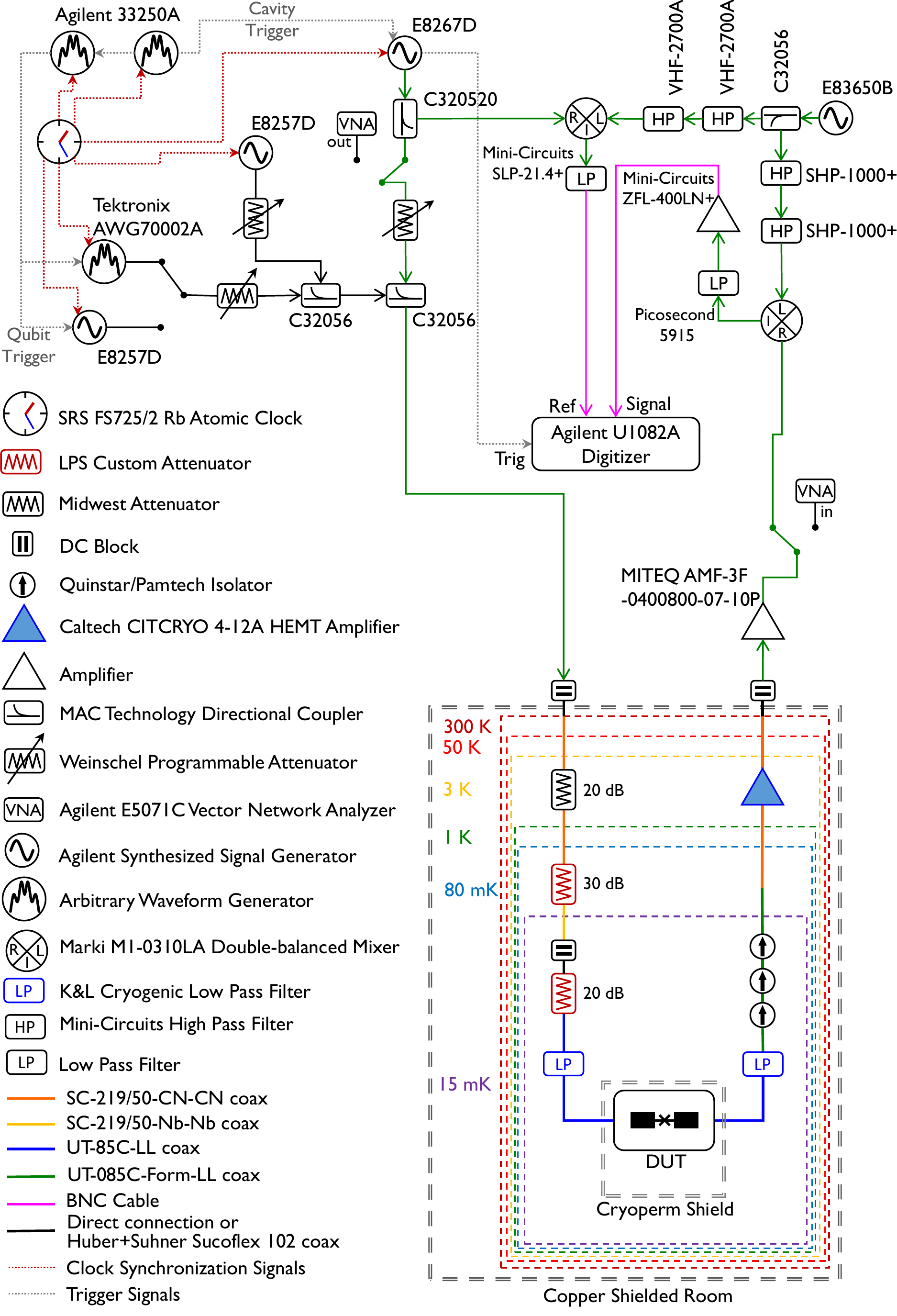}
\caption{Schematic of the complete control and measurement network}
\end{figure}

\vfill
\hfill
\clearpage
\newpage

\section{Calibration for \textsf{SWIPHT} Pulse}

\begin{figure}[h]
\centering
\includegraphics[width=1\columnwidth]{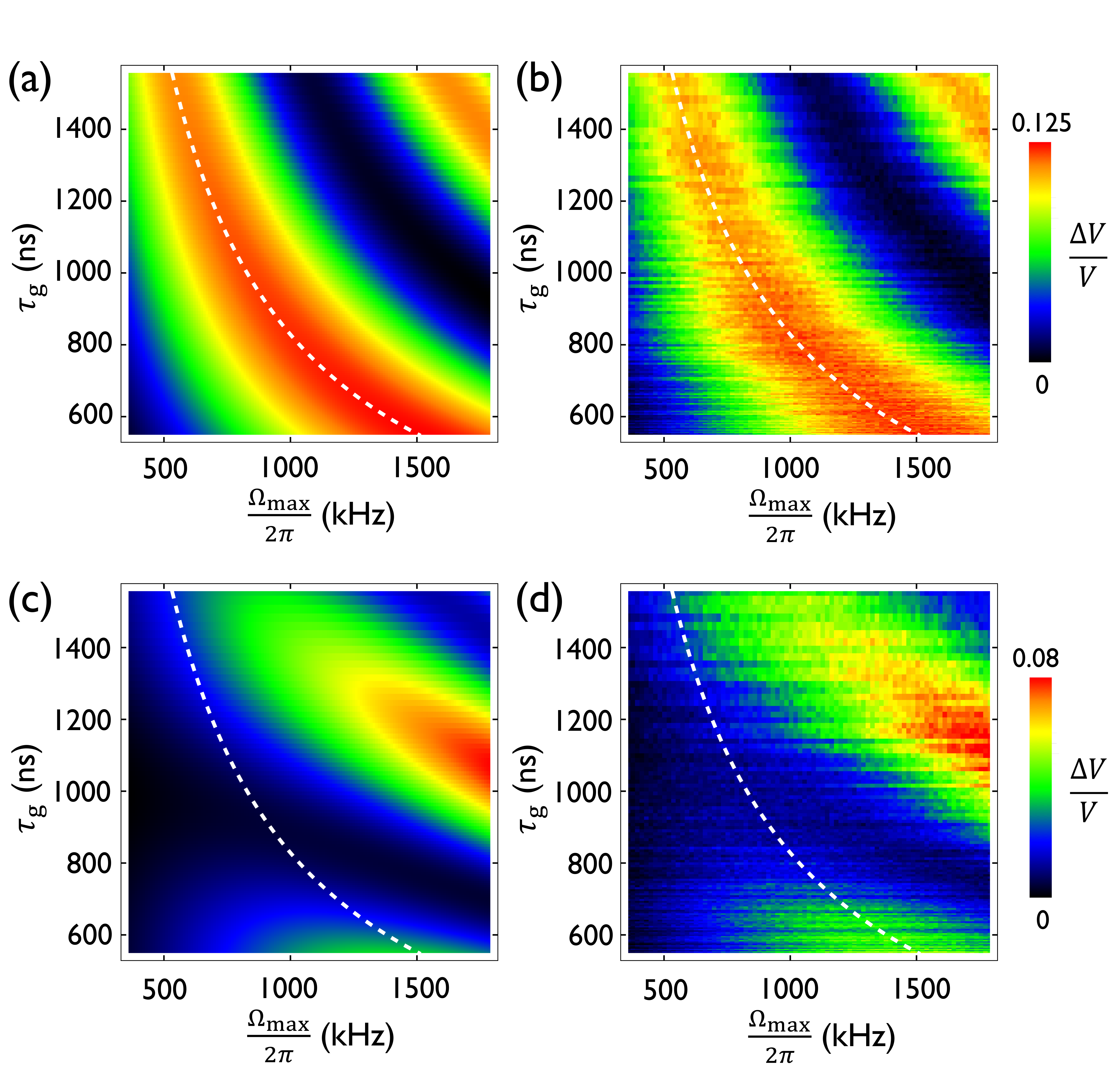}
\caption{Gate time ($\tau_\textnormal{g}$) and maximum pulse amplitude ($\Omega_\textnormal{max}$) calibration results for the \textsf{SWIPHT} pulse. (a) Simulated and (b) Experimental qubit H final population with control qubit on. (c) Simulated and (d) Experimental qubit H final population with control qubit off. White dashed lines depict the theoretical relationship between $\tau_\textnormal{g}$ and $\Omega_\textnormal{max}$. \label{fig:SWIPHT_calib}}
\end{figure}

Initial calibration for the \textsf{SWIPHT} pulse was performed by generating the theoretical \textsf{SWIPHT} pulse shape with varying $\tau_\textnormal{g}$ and $\Omega_\textnormal{max}$. The readout was biased at a point both sensitive to the state of qubit H and relatively insensitive to the state of qubit L. The ability for the pulse to excite qubit H controlled on the state of qubit L was evaluated over the 2D landscape. Good agreement was obtained between the master equation simulations (Fig.~\ref{fig:SWIPHT_calib}(a) and Fig.~\ref{fig:SWIPHT_calib}(c)) and experiments (Fig.~\ref{fig:SWIPHT_calib}(b) and Fig.~\ref{fig:SWIPHT_calib}(d)). We confirmed that $\Omega_\textnormal{max}/2\pi = \SI{913}{\kHz}$ and $\tau_\textnormal{g} = \SI{907}{\ns}$ corresponding to $\chi_\textnormal{qq}/2\pi = \SI{0.52}{\MHz}$ was the optimal set of parameters.

\vfill
\hfill
\clearpage
\newpage

\section{QPT process matrices for SWIPHT gates}

The comparisons between real and imaginary parts of $\chi$-matrices obtained from the experiment, master-equation simulations and decoherence-free ideal gates when implementing the SWIPHT protocol are shown below. Either qubit was used as the control and two values of $\phi_\textrm{d}$ were used for each case. The maximum theoretical magnitude for any element is 0.25.


\begin{figure}[ht]
\centering
\includegraphics[width=1\columnwidth]{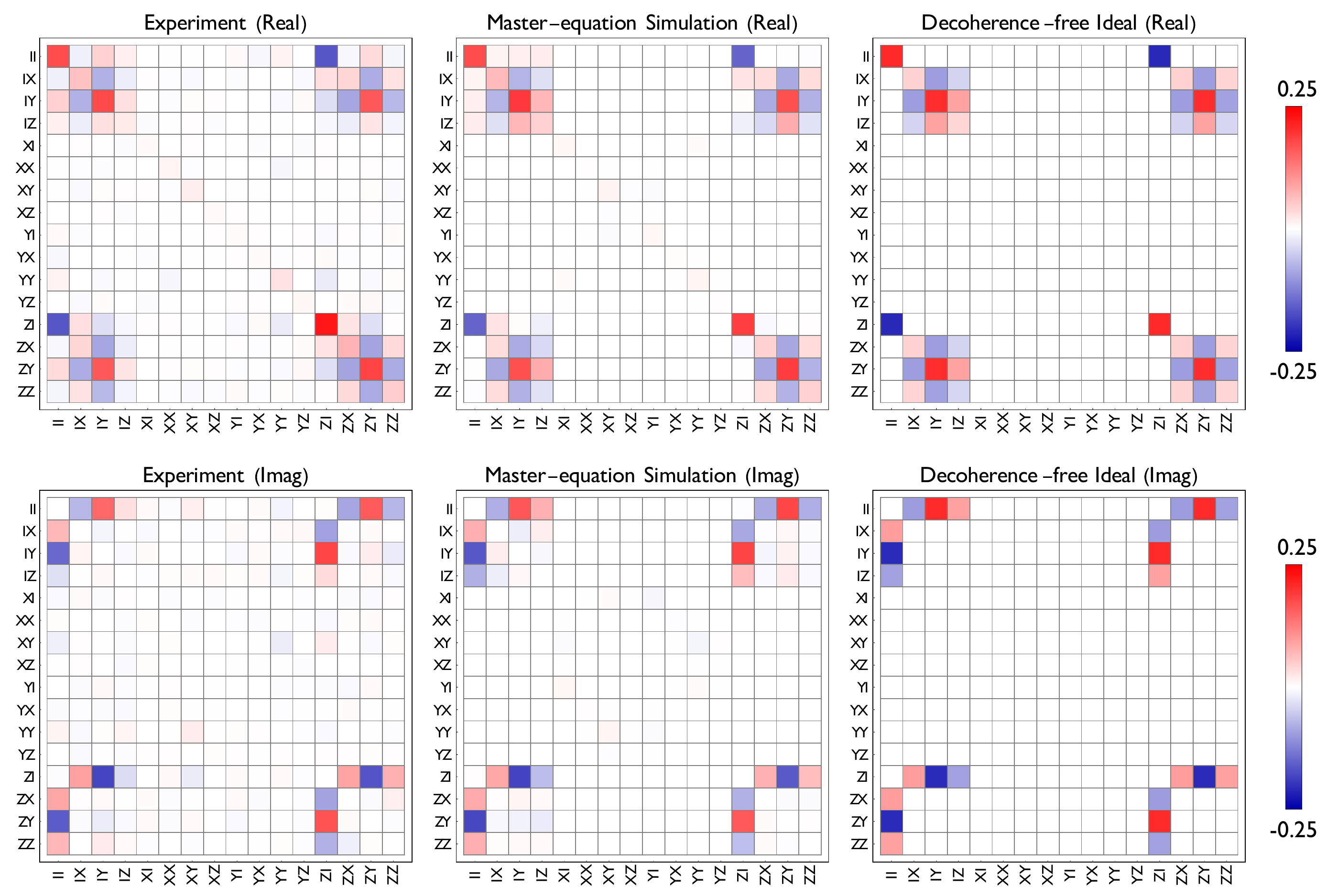}
\caption{$\chi$-matrix for $\textsf{SWIPHT}^{49\pi/36}_{\textrm{L}}$. Qubit L as the control and $\phi_\textrm{d} = 49\pi/36$. }
\end{figure}

\begin{figure}[ht]
\centering
\includegraphics[width=1\columnwidth]{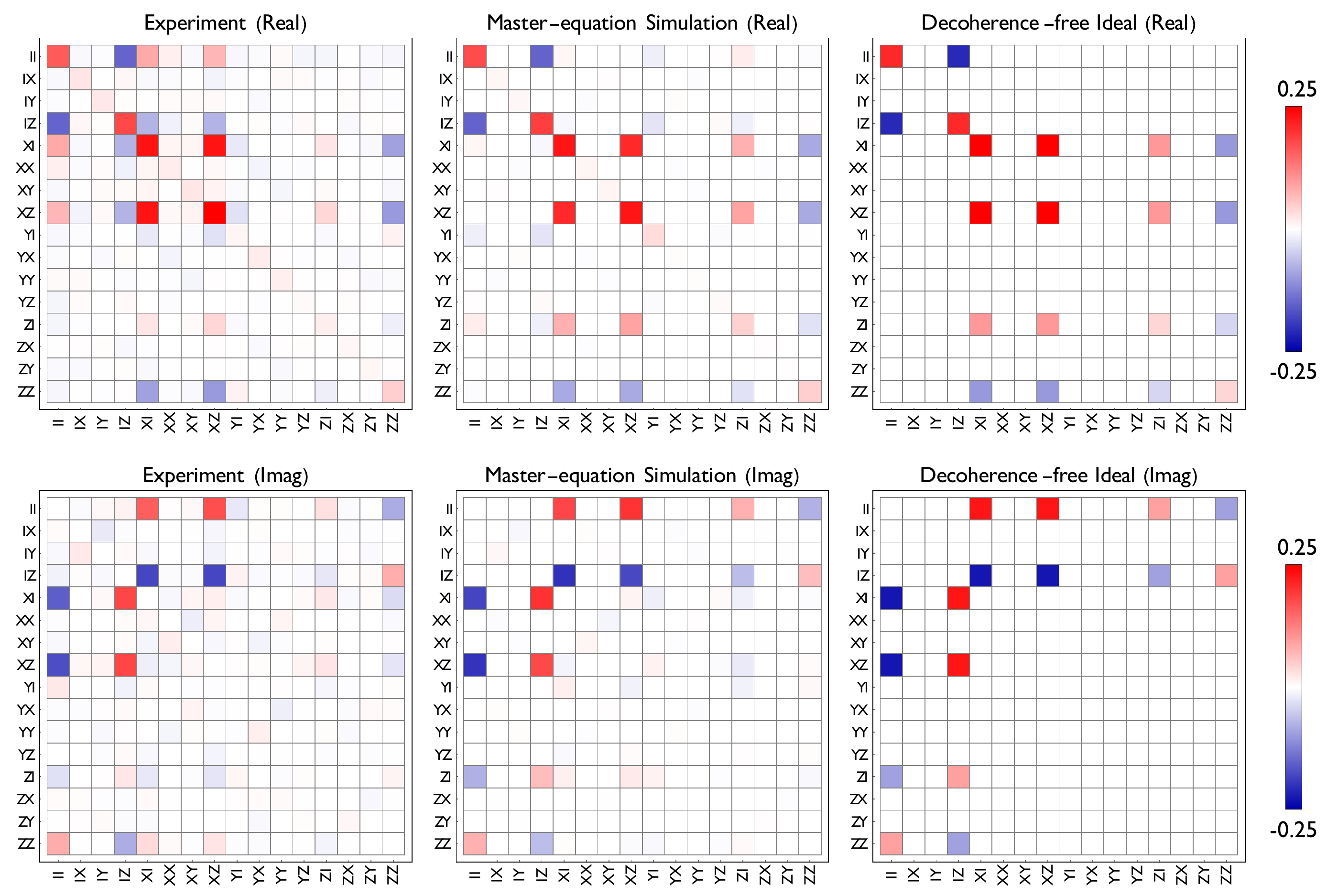}
\caption{$\chi$-matrix for $\textsf{SWIPHT}^0_{\textrm{H}}$. Qubit H as the control and $\phi_\textrm{d} = 0$.}
\end{figure}

\begin{figure}[ht]
\centering
\includegraphics[width=1\columnwidth]{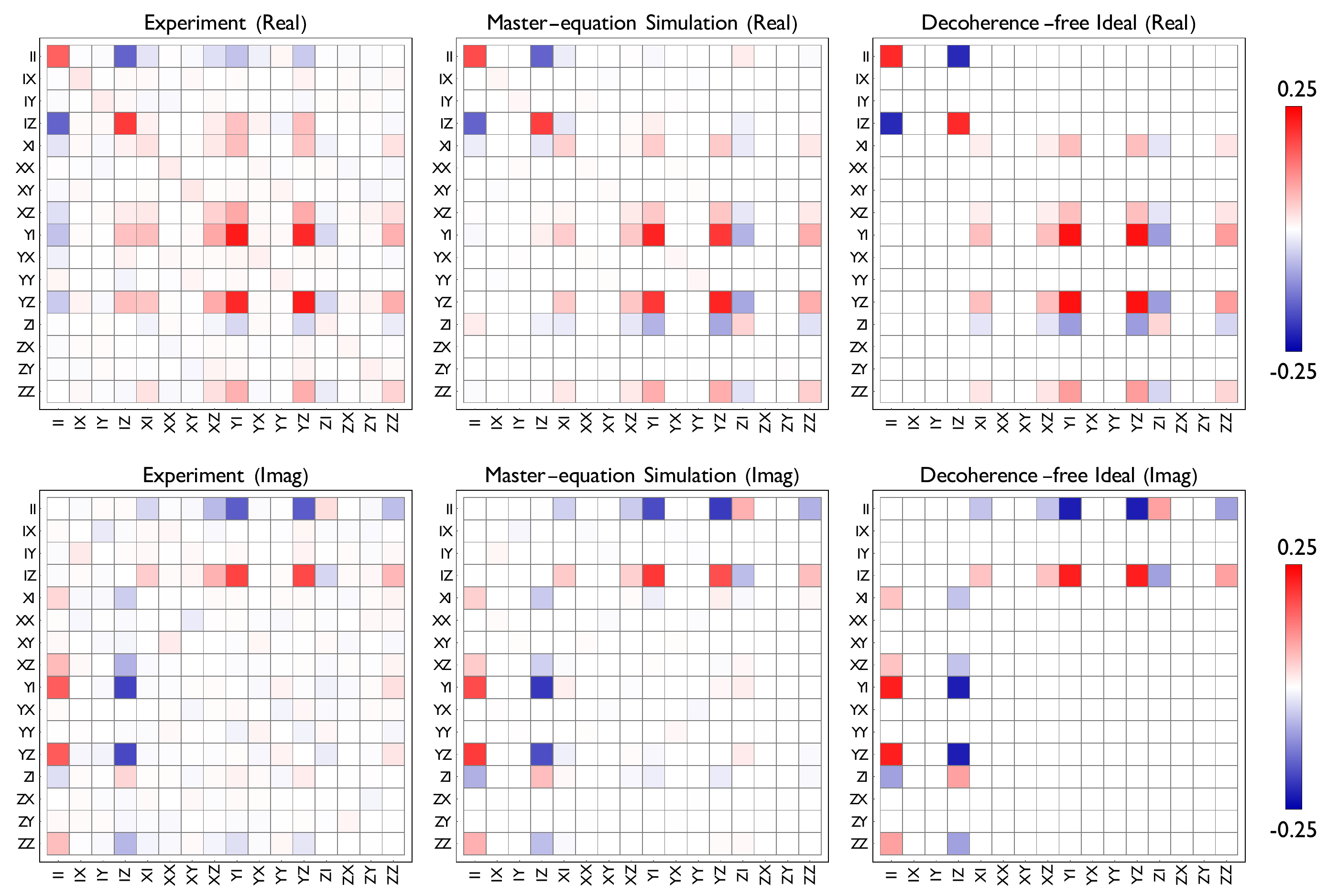}
\caption{$\chi$-matrix for $\textsf{SWIPHT}^{7\pi/12}_{\textrm{H}}$. Qubit H as the control and $\phi_\textrm{d} = 7\pi/12$.}
\end{figure}

\vfill
\hfill
\clearpage
\newpage

\section{QPT process matrices for single-qubit gates}

The comparisons between real and imaginary parts of $\chi$-matrices obtained from the experiment, and decoherence-free ideal gates when implementing the single-qubit gates are shown below. The maximum theoretical magnitude for any element is $1.0$ for $R^\pi$ gates and $0.5$ for $R^{\pi/2}$ gates.

\begin{figure}[ht]
\centering
\includegraphics[width=1\columnwidth]{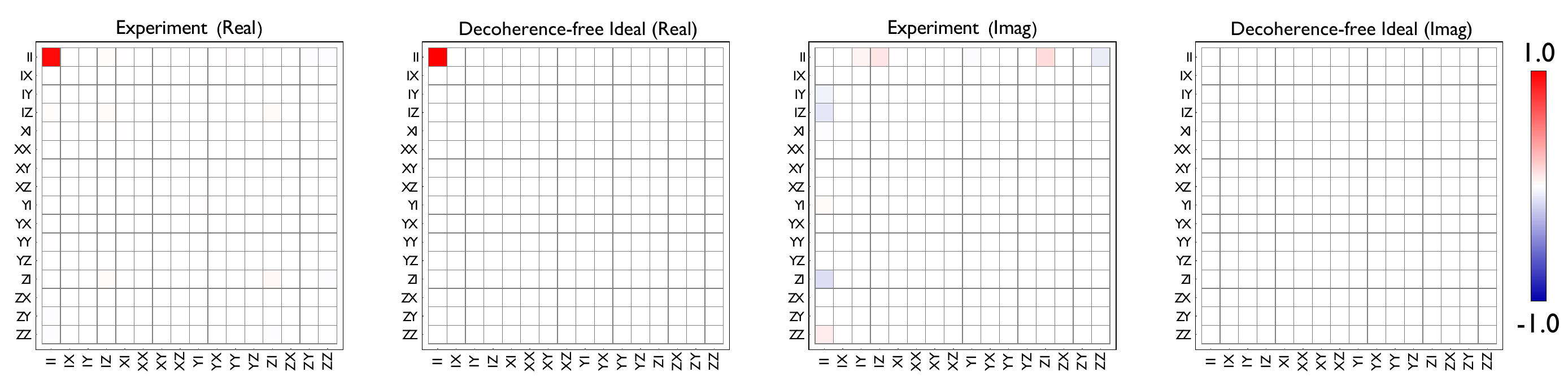}
\caption{$\chi$-matrix for $I \otimes I$.}
\end{figure}

\begin{figure}[ht]
\centering
\includegraphics[width=1\columnwidth]{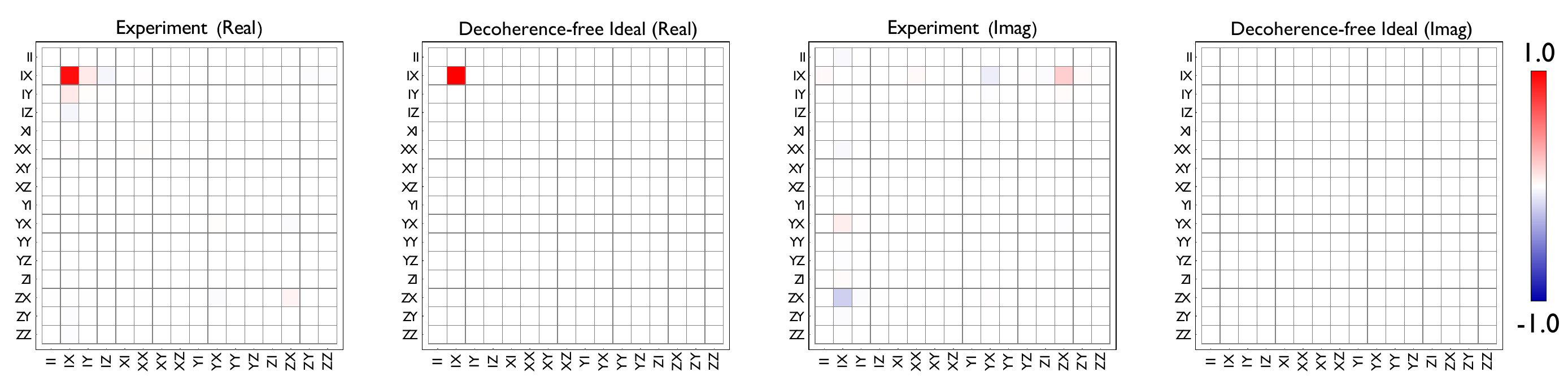}
\caption{$\chi$-matrix for $I \otimes R_x^\pi$.}
\end{figure}

\begin{figure}[ht]
\centering
\includegraphics[width=1\columnwidth]{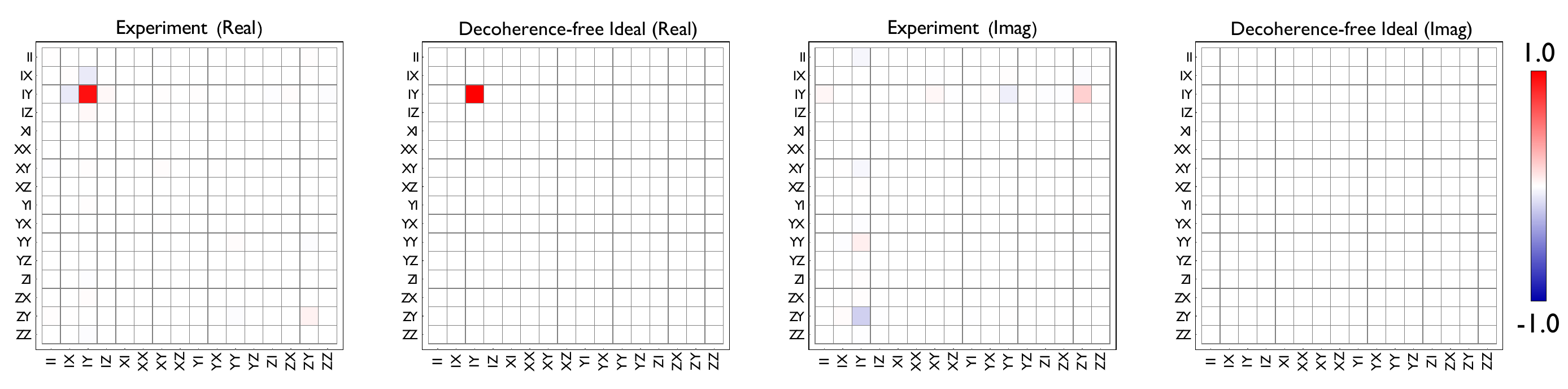}
\caption{$\chi$-matrix for $I \otimes R_y^\pi$.}
\end{figure}

\begin{figure}[ht]
\centering
\includegraphics[width=1\columnwidth]{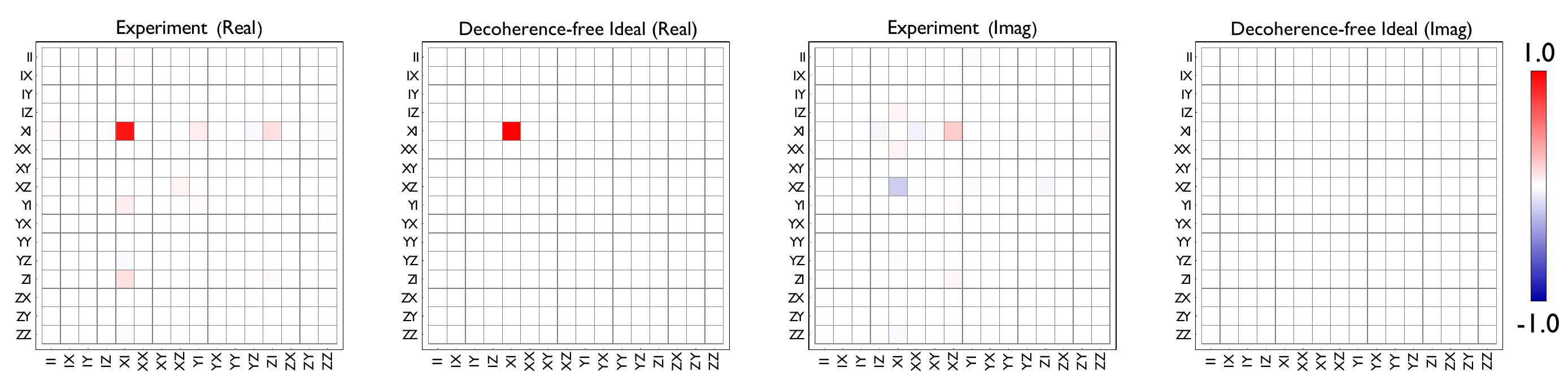}
\caption{$\chi$-matrix for $R_x^\pi \otimes I$.}
\end{figure}

\begin{figure}[ht]
\centering
\includegraphics[width=1\columnwidth]{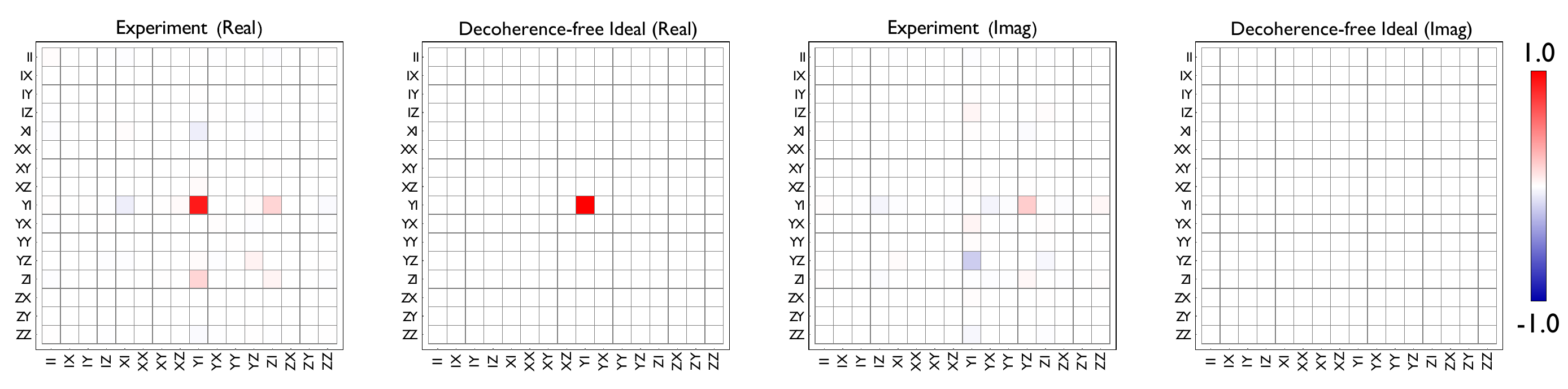}
\caption{$\chi$-matrix for $R_y^\pi \otimes I$.}
\end{figure}

\begin{figure}[ht]
\centering
\includegraphics[width=1\columnwidth]{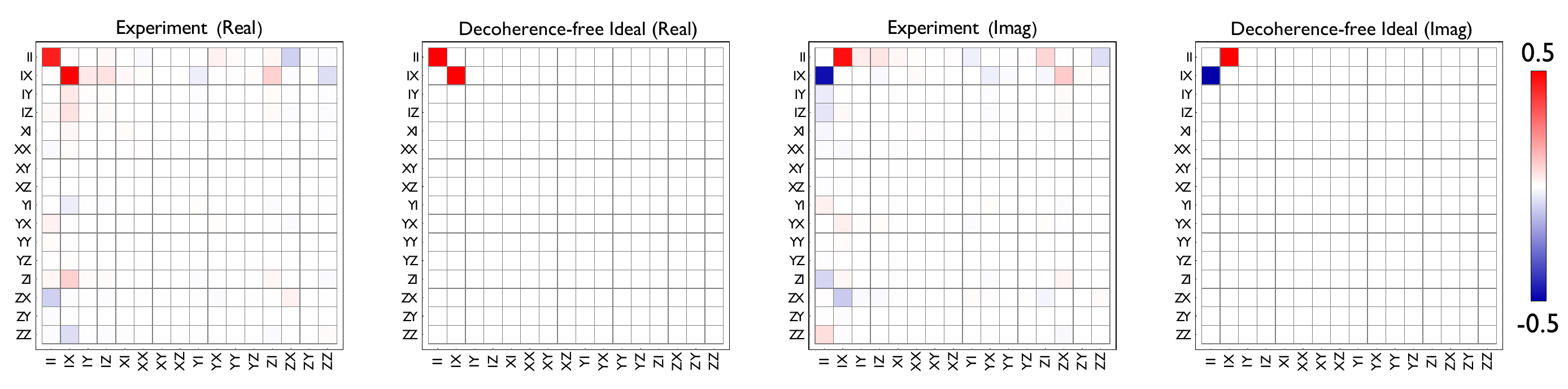}
\caption{$\chi$-matrix for $I \otimes R_x^{\pi/2}$.}
\end{figure}

\begin{figure}[ht]
\centering
\includegraphics[width=1\columnwidth]{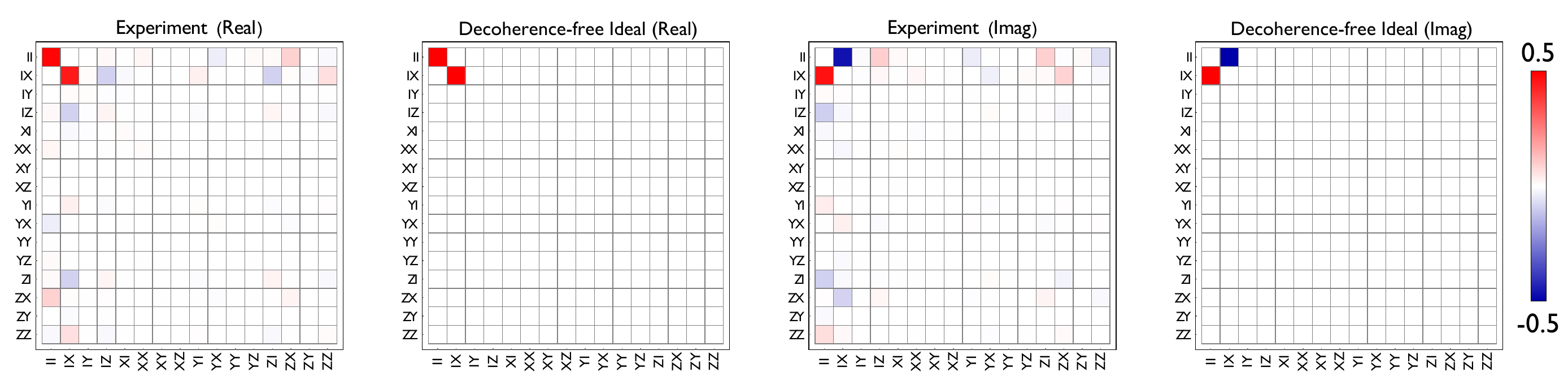}
\caption{$\chi$-matrix for $I \otimes R_x^{-\pi/2}$.}
\end{figure}

\begin{figure}[ht]
\centering
\includegraphics[width=1\columnwidth]{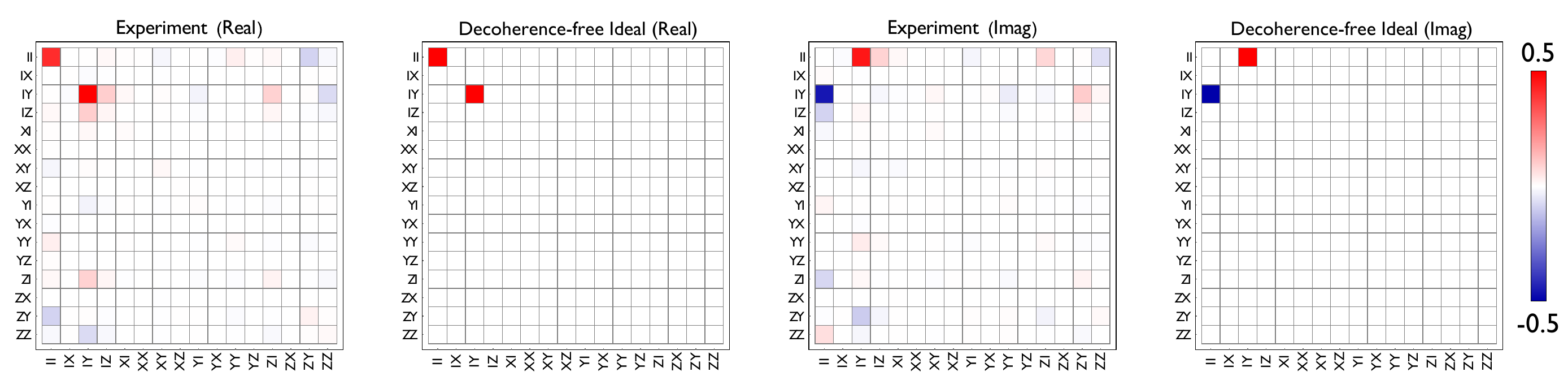}
\caption{$\chi$-matrix for $I \otimes R_y^{\pi/2}$.}
\end{figure}

\begin{figure}[ht]
\centering
\includegraphics[width=1\columnwidth]{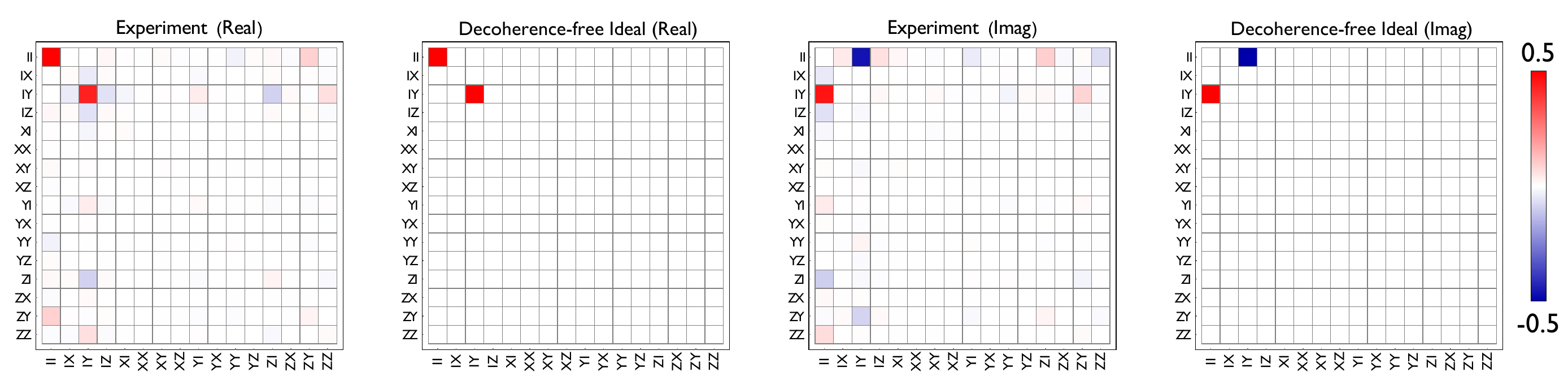}
\caption{$\chi$-matrix for $I \otimes R_y^{-\pi/2}$.}
\end{figure}

\begin{figure}[ht]
\centering
\includegraphics[width=1\columnwidth]{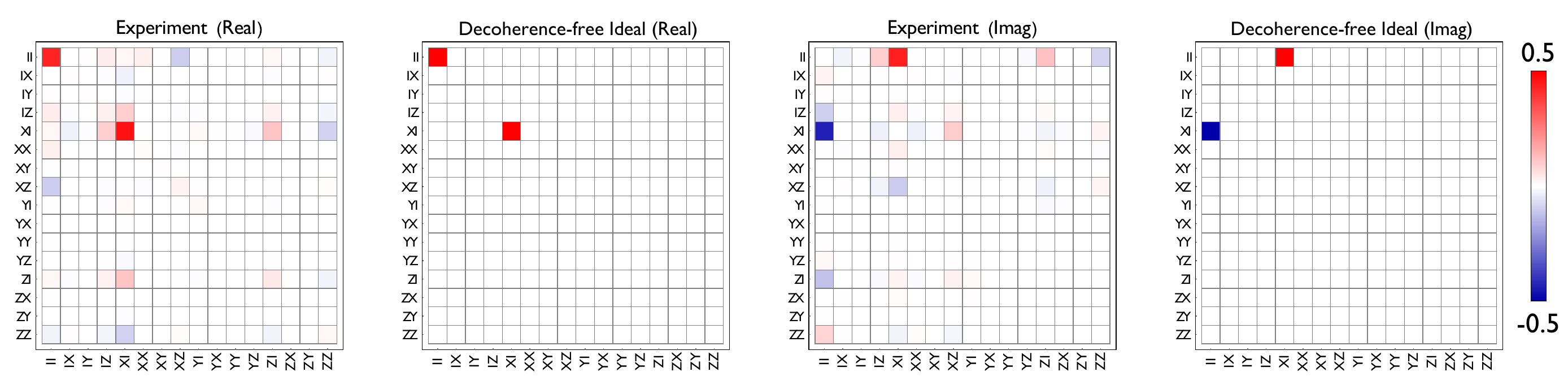}
\caption{$\chi$-matrix for $R_x^{\pi/2} \otimes I$.}
\end{figure}

\begin{figure}[ht]
\centering
\includegraphics[width=1\columnwidth]{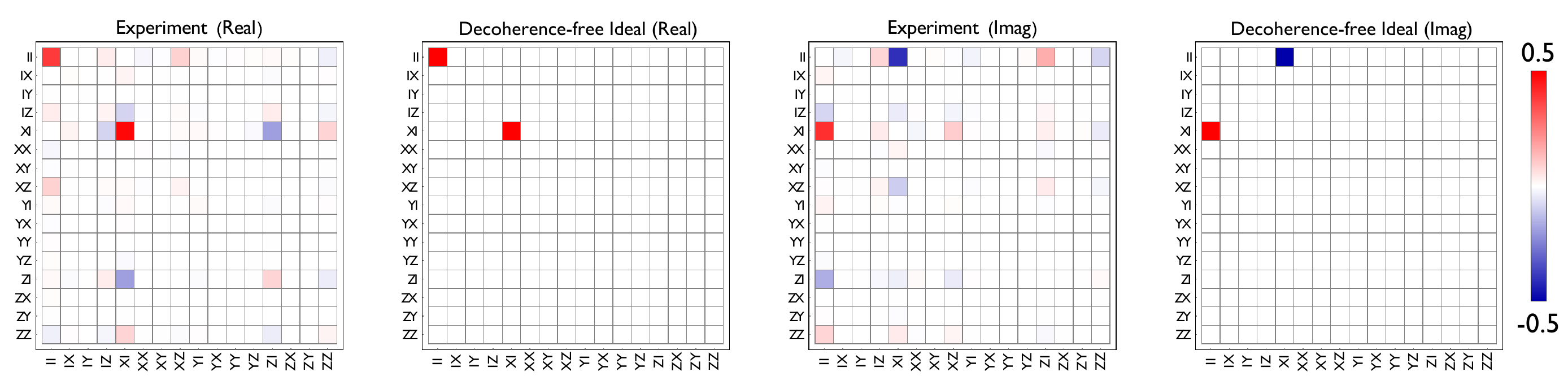}
\caption{$\chi$-matrix for $R_x^{-\pi/2} \otimes I$.}
\end{figure}

\begin{figure}[ht]
\centering
\includegraphics[width=1\columnwidth]{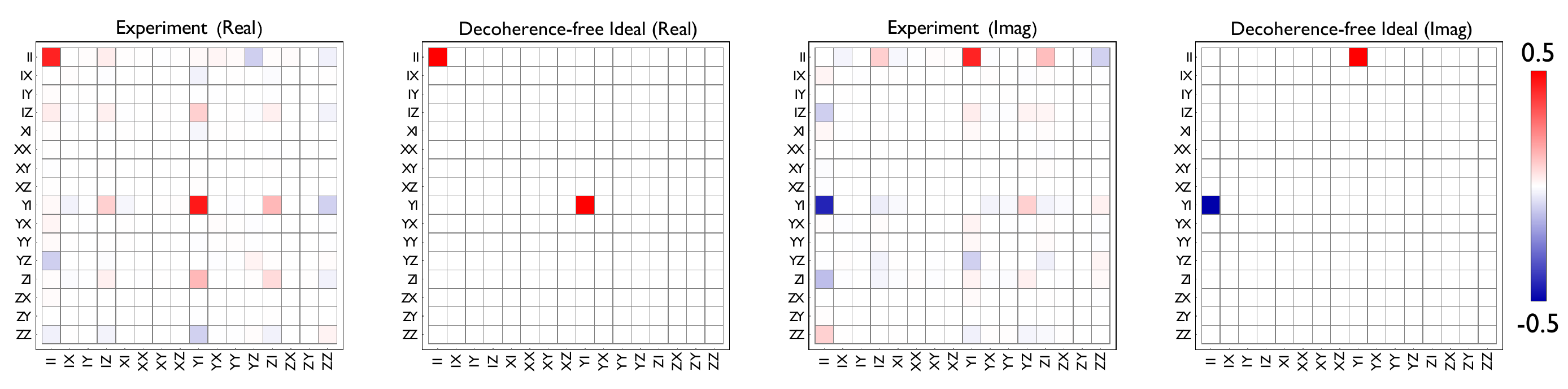}
\caption{$\chi$-matrix for $R_y^{\pi/2} \otimes I$.}
\end{figure}

\begin{figure}[h]
\centering
\includegraphics[width=1\columnwidth]{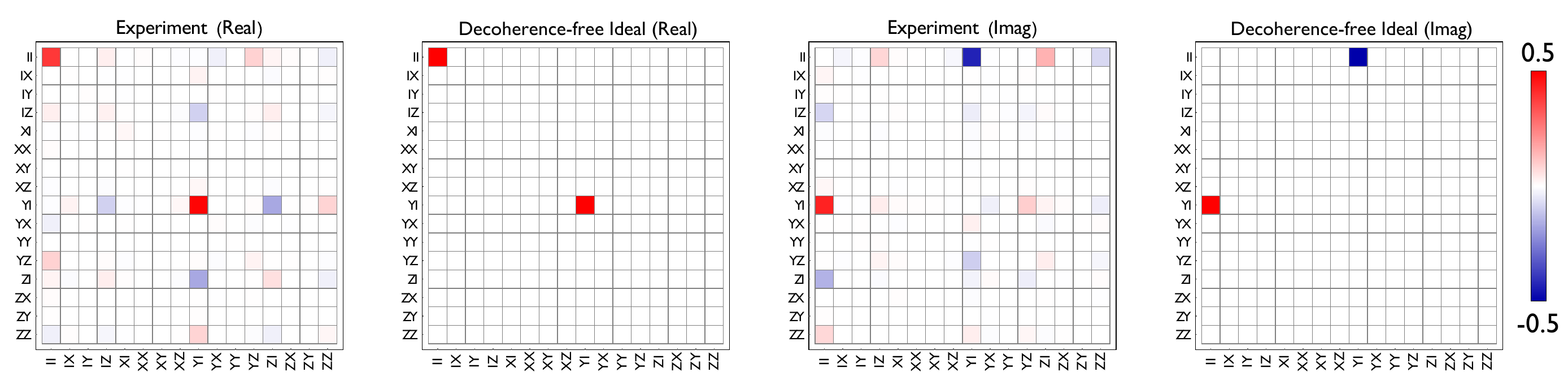}
\caption{$\chi$-matrix for $R_y^{-\pi/2} \otimes I$.}
\end{figure}